%% file: main.tex
\documentclass[reprint,onecolumn,notitlepage,preprintnumbers,nofootinbib,amsmath,amssymb,aps,prd,floatfix,superscriptaddress,longbibliography,svgnames]{revtex4-1}

\usepackage{amsmath}
\usepackage{amsfonts}
\usepackage{amssymb}
\usepackage{graphicx}
\usepackage{bm}
\usepackage{subfigure}
\usepackage{url}
\usepackage[hyperindex]{hyperref}
\usepackage{color}
\usepackage[ddmmyy,24hr]{datetime}
\usepackage{bigdelim}
\usepackage{booktabs}
\usepackage{dcolumn}
\usepackage{multirow}
\usepackage{subfigure}
\usepackage{cancel}
\usepackage{stackrel}
\usepackage{paralist}
\usepackage{xspace}
\usepackage{slashed}
\usepackage{cancel}
\usepackage{todonotes}
\usepackage{enumerate}
\usepackage{float}
\usepackage{fullpage}
\newcommand{\cenns}{CE$\nu$NS\xspace}

\begin{document}

\title{Probing light mediators and $(g-2)_{\mu}$ through detection of coherent elastic neutrino nucleus scattering at COHERENT}

\author{M. Atzori Corona}
\affiliation{Dipartimento di Fisica, Universit\`{a} degli Studi di Cagliari,
	Complesso Universitario di Monserrato - S.P. per Sestu Km 0.700,
	09042 Monserrato (Cagliari), Italy}
\affiliation{Istituto Nazionale di Fisica Nucleare (INFN), Sezione di Cagliari,
	Complesso Universitario di Monserrato - S.P. per Sestu Km 0.700,
	09042 Monserrato (Cagliari), Italy}

\author{M. Cadeddu}
\affiliation{Istituto Nazionale di Fisica Nucleare (INFN), Sezione di Cagliari,
	Complesso Universitario di Monserrato - S.P. per Sestu Km 0.700,
	09042 Monserrato (Cagliari), Italy}

\author{N. Cargioli}
\affiliation{Dipartimento di Fisica, Universit\`{a} degli Studi di Cagliari,
	Complesso Universitario di Monserrato - S.P. per Sestu Km 0.700,
	09042 Monserrato (Cagliari), Italy}
\affiliation{Istituto Nazionale di Fisica Nucleare (INFN), Sezione di Cagliari,
	Complesso Universitario di Monserrato - S.P. per Sestu Km 0.700,
	09042 Monserrato (Cagliari), Italy}

\author{F. Dordei}
\affiliation{Istituto Nazionale di Fisica Nucleare (INFN), Sezione di Cagliari,
	Complesso Universitario di Monserrato - S.P. per Sestu Km 0.700,
	09042 Monserrato (Cagliari), Italy}

\author{C. Giunti}
\affiliation{Istituto Nazionale di Fisica Nucleare (INFN), Sezione di Torino, Via P. Giuria 1, I--10125 Torino, Italy}

\author{Y.F. Li}
\affiliation{Institute of High Energy Physics,
	Chinese Academy of Sciences, Beijing 100049, China}
\affiliation{School of Physical Sciences, University of Chinese Academy of Sciences, Beijing 100049, China}

\author{E. Picciau}
\affiliation{Dipartimento di Fisica, Universit\`{a} degli Studi di Cagliari,
	Complesso Universitario di Monserrato - S.P. per Sestu Km 0.700,
	09042 Monserrato (Cagliari), Italy}
\affiliation{Istituto Nazionale di Fisica Nucleare (INFN), Sezione di Cagliari,
	Complesso Universitario di Monserrato - S.P. per Sestu Km 0.700,
	09042 Monserrato (Cagliari), Italy}

\author{C. A. Ternes}
\affiliation{Istituto Nazionale di Fisica Nucleare (INFN), Sezione di Torino, Via P. Giuria 1, I--10125 Torino, Italy}

\author{Y.Y. Zhang}
\email{zhangyiyu@ihep.ac.cn}
\affiliation{Institute of High Energy Physics,
	Chinese Academy of Sciences, Beijing 100049, China}
\affiliation{School of Physical Sciences, University of Chinese Academy of Sciences, Beijing 100049, China}

\date{\dayofweekname{\day}{\month}{\year} \ddmmyydate\today, \currenttime}

\begin{abstract}
We present the constraints on the parameters of several light boson mediator models
obtained from the analysis of the current data of the COHERENT CE$\nu$NS experiment.
We consider a variety of vector boson mediator models:
the so-called universal,
the $B-L$ and other
anomaly-free $U(1)'$ gauge models with direct couplings of the new vector boson with neutrinos and quarks,
and the anomaly-free
$L_e-L_\mu$,
$L_e-L_\tau$, and
$L_\mu-L_\tau$
gauge models where the coupling of the new vector boson with the quarks
is generated by kinetic mixing with the photon at the one-loop level.
We consider also a model with a new light scalar boson mediator
that is assumed, for simplicity, to have universal coupling
with quarks and leptons.
Since the COHERENT CE$\nu$NS data
are well-fitted with the cross section predicted by the Standard Model,
the analysis of the data yields constraints for the mass and coupling of the new boson mediator that depend on the charges of quarks and neutrinos
in each model under consideration.
We compare these constraints
with the limits obtained in other experiments
and with the values that can explain the muon $g-2$ anomaly
in the models where the muon couples to the new boson mediator.
\end{abstract}

\maketitle

\section{Introduction}

The discovery of coherent elastic neutrino-nucleus scattering (\cenns) in cesium-iodide (CsI) by the COHERENT Collaboration~\cite{Akimov:2017ade,Akimov:2018vzs} sparked a flood of research into a variety of physical processes, with substantial implications for particle physics, astrophysics, nuclear physics, and beyond~\cite{Cadeddu:2017etk,Papoulias:2019lfi,Coloma:2017ncl,Liao:2017uzy,Lindner:2016wff,Giunti:2019xpr,Denton:2018xmq,AristizabalSierra:2018eqm,Cadeddu:2018dux,Papoulias:2017qdn,Cadeddu:2019eta,Papoulias:2019txv,Khan:2019cvi,Dutta:2019eml,AristizabalSierra:2018eqm,Cadeddu:2018izq,Dutta:2019nbn,Abdullah:2018ykz,Ge:2017mcq}.
After the discovery in 2017 of \cenns with the CsI detector,
the COHERENT Collaboration
accomplished in 2020 the first observation of \cenns in argon (Ar)~\cite{COHERENT:2020iec,COHERENT:2020ybo}
and
updated in 2021 the results obtained with the CsI detector~\cite{Akimov:2021dab}.
By combining the greater \cenns statistics with a refined quenching factor estimation for the CsI measurement and by virtue of the complementary role of two different target nuclei, more stringent tests of nuclear physics, neutrino properties, electroweak interactions, and new physics beyond the Standard Model (SM) have been performed~\cite{Cadeddu:2020lky,Miranda:2020tif,Cadeddu:2020nbr,Cadeddu:2021ijh,Banerjee:2021laz}.

The \cenns process happens when the momentum transfer between the incoming neutrino and the target nucleus is so small that the wavelength of the boson which mediates the interaction is larger than the nuclear radius, so that the neutrino interacts with the nucleus as a whole and the cross section is proportional to the square of the number of nucleons participating to the process.
The \cenns process is a pure neutral current interaction which is mediated by the exchange of the $Z$ vector boson in the SM, but it can also receive contributions from other hypothetical neutral bosons in theories beyond the SM. Therefore, it turns out to be a powerful tool to probe new physics interactions beyond the SM~\cite{Coloma:2017ncl,Liao:2017uzy,Lindner:2016wff,Giunti:2019xpr,Denton:2018xmq,AristizabalSierra:2018eqm}.

In this paper we test new physics models with interactions mediated by
a light vector or scalar boson that contribute to the \cenns process
by analyzing the recently released 2021 CsI data~\cite{Akimov:2021dab}
and the 2020 Ar data~\cite{COHERENT:2020iec,COHERENT:2020ybo}
of the COHERENT experiment.
For each model,
we present the constraints on the mass and coupling of the light vector or scalar boson mediator
that we obtained from the separate and combined fits of the CsI and Ar COHERENT \cenns data.
Comparing with the previous publication in Ref.~\cite{Cadeddu:2020nbr},
we have considered a larger variety of vector mediator models, we have included the scalar mediator model,
and we have updated the analysis using the recently released 2021 CsI data~\cite{Akimov:2021dab}.

We also consider the possible explanation of
the $4.2\sigma$ difference between the SM prediction~\cite{Aoyama:2020ynm,Aoyama:2012wk,Aoyama:2019ryr,Czarnecki:2002nt,Gnendiger:2013pva,Davier:2017zfy,Keshavarzi:2018mgv,Colangelo:2018mtw,Hoferichter:2019gzf,Davier:2019can,Keshavarzi:2019abf,Melnikov:2003xd,Masjuan:2017tvw,Colangelo:2017fiz,Hoferichter:2018kwz,Gerardin:2019vio,Bijnens:2019ghy,Colangelo:2019uex,Pauk:2014rta,Danilkin:2016hnh,Jegerlehner:2017gek,Knecht:2018sci,Eichmann:2019bqf,Roig:2019reh,Blum:2019ugy,Colangelo:2014qya}
of the value of the muon anomalous magnetic moment $(g-2)_\mu$
and the combination of the values measured at
the Brookhaven National Laboratory~\cite{Muong-2:2006rrc}
and recently at the Fermi National Laboratory~\cite{Muong-2:2021ojo}.
This so-called $(g-2)_\mu$ anomaly is a putative signal of physics beyond the SM,
which has been studied in many papers
(see, e.g., Refs.~\cite{Athron:2021iuf,Lindner:2016bgg}).
Interestingly,
several light vector mediator models are regarded as candidate solutions
(see, e.g., Refs.~\cite{Baek:2001kca,Ma:2001md,Altmannshofer:2016oaq,Amaral:2021rzw,Cadeddu:2021dqx,Zhou:2021vnf,Ko:2021lpx,Hapitas:2021ilr,Cheng:2021okr}),
but also a light scalar mediator have the potential to solve the anomaly~\cite{Jegerlehner:2009ry}.
Among the models that we consider in this paper,
those in which the muon interacts with the
new light boson mediator can explain the $(g-2)_\mu$ anomaly.
For these models, we compare the constraints
on the mass and coupling of the new light boson mediator obtained from the analysis of the COHERENT \cenns data
and those obtained by other experiments focusing on the parameter region that can solve the $(g-2)_\mu$ anomaly.
There are many studies of extensions of the SM with the addition of a $U(1)'$ gauge group with an associated neutral vector gauge boson $Z'$
(see, e.g., the review in Ref.~\cite{Langacker:2008yv}).
The models differ in the charges of the fermions,
which determine the contributions to \cenns of the interactions mediated by
the $Z'$ vector boson.
These contributions
add coherently to the SM weak neutral current interactions which are mediated by the $Z$ vector boson.
The effects are quantified by additional terms in the weak charge of the nucleus.
Note that the effects on the \cenns process
of the interactions mediated by a light boson are different from those
induced by the so-called non-standard interactions (NSI),
that arise in an effective four-fermion theory in which the heavy mediator has been integrated out.
In the case of NSI there is a global rescaling of the \cenns cross section that depends on the interaction parameters of the NSI,
whereas a light boson mediator can
alter the nuclear recoil energy spectrum through the boson propagator
that depends on the momentum transfer.
This effect generates distinct spectral features that can be probed
with the experimental observations.

In this paper we first consider the so-called universal $Z'$ model in which all the standard fermions have the same
charge~\cite{Liao:2017uzy,Papoulias:2017qdn,Billard:2018jnl,Papoulias:2019txv,Khan:2019cvi,Cadeddu:2020nbr,Bertuzzo:2021opb}.
This model is not anomaly-free per se, but it can be extended with new non-standard particles to make it anomaly-free.
Then, we consider several $U(1)'$ models in which quarks and leptons have appropriate non-zero charges that cancel the quantum anomalies
(e.g., the popular $B-L$ model~\cite{Langacker:2008yv,Mohapatra:2014yla,Okada:2018ktp},
where $B$ is the baryon number and $L$ is the total lepton number).
Since in these models the $Z'$ vector boson interacts directly with neutrinos and nucleons,
the \cenns process occurs at tree level and it is possible to obtain
stringent constraints on the mass and coupling of the new vector boson
from the COHERENT \cenns data.

We also consider the anomaly-free
$L_e-L_\mu$,
$L_e-L_\tau$, and
$L_\mu-L_\tau$
$U(1)'$ models~\cite{Foot:1990mn,Foot:1990uf,He:1990pn,Foot:1992ui}
(where $L_{\alpha}$ are the lepton generation numbers, for $\alpha=e, \mu, \tau$)
in which the charges are exclusively leptonic.
However, in these models there are contributions to the \cenns process,
which occur through the kinetic mixing of the $Z'$ boson with the photon,
that is generated at one-loop
level~\cite{Altmannshofer:2019zhy,Banerjee:2018mnw,Banerjee:2021laz},
and the interaction of the photon with the protons in the target nuclei.
Therefore,
we can constrain the mass and coupling of the vector boson in these models
using the COHERENT \cenns data,
albeit less tightly than in the models with direct quark-$Z'$ interactions,
because of the weaker one-loop interaction.

We consider also contributions to the \cenns process
of interactions mediated by a light scalar boson~\cite{Lindner:2016wff,Cerdeno:2016sfi,Farzan:2018gtr,AristizabalSierra:2018eqm,AristizabalSierra:2019ykk},
which differ from those mediated by a light vector boson for the following two fundamental reasons.
First, the helicity-flipping interactions mediated by a scalar boson contribute incoherently to the \cenns process with respect to the helicity-conserving SM contribution,
contrary to the helicity-conserving interactions mediated by a new vector boson
that contribute coherently.
Therefore, in the scalar case, the new contribution consists in an addition to the cross section,
not to the amplitude of the process as in the vector case.
Second,
the scalar charges of the nucleons are not simply given by the sum of the charges of the valence quarks as in the vector case,
because the scalar currents are not conserved as the vector currents.
Therefore, the scalar charges of the nucleons must be calculated and the results have large theoretical uncertainties.

The paper is organized as follows.
In Section~\ref{sec:method} the method of the COHERENT data analysis is described.
In Section~\ref{sec:cs}, we present the cross section of the \cenns process and
summarize the models of light mediators and the corresponding effects on the \cenns cross section. 
In Section~\ref{sec:result}, the COHERENT \cenns constraints on the allowed parameter space of the light mediator models are presented and compared with the $(g-2)_\mu$ allowed regions and other current limits.
Finally, we conclude and summarize our results in Section~\ref{sec:conclusions}.

\section{COHERENT data analysis}
\label{sec:method}

The \cenns event energy spectra in the COHERENT experiment depend on the neutrino flux produced by the pion decay.
The total differential neutrino flux is given by the sum of the three neutrino components, where the first prompt component is coming from the pion decay
($\pi^{+} \rightarrow \mu^{+}+\nu_{\mu}$), and the second two delayed components are coming from the subsequent muon decay ($\mu^{+} \rightarrow e^{+}+\nu_{e}+\bar{\nu}_{\mu}$)
\begin{align}
	\frac{d N_{\nu_{\mu}}}{d E}&=\eta \, \delta\!\left(E-\frac{m_{\pi}^{2}-m_{\mu}^{2}}{2 m_{\pi}}\right), \\
	\frac{d N_{\nu_{\tilde{\mu}}}}{d E}&=\eta \, \frac{64 E^{2}}{m_{\mu}^{3}}\left(\frac{3}{4}-\frac{E}{m_{\mu}}\right), \\
	\frac{d N_{\nu_{e}}}{d E}&=\eta \, \frac{192 E^{2}}{m_{\mu}^{3}}\left(\frac{1}{2}-\frac{E}{m_{\mu}}\right).
\end{align}
Here, $E$ is the neutrino energy, $m_{\pi}$ and $m_{\mu}$ are the pion and muon masses, and $\eta=r N_{\mathrm{POT}} / 4 \pi L^{2}$ is the normalization factor, 
where $r$ is the number of neutrinos per flavor produced for each proton-on-target (POT), 
$N_{\mathrm{POT}}$ is the number of POT, 
and $L$ is the baseline between the source and the detector.
For the COHERENT Ar detector, called CENNS-10, we use
$r=0.09$, $N_{\mathrm{POT}}=13.7 \cdot 10^{22}$ and $L=27.5~\mathrm{m}$~\cite{COHERENT:2020ybo}.
For the COHERENT CsI detector, we use
$r=0.0848$, $N_{\mathrm{POT}}=3.198 \cdot 10^{23}$ and $L=19.3~\mathrm{m}$~\cite{Akimov:2021dab}.

The theoretical \cenns event number $N^\mathrm{CE \nu NS}_{i}$ in each nuclear-recoil energy-bin $i$ is given by
\begin{equation}\label{N_cevns}
N_{i}^{\mathrm{CE}\nu\mathrm{NS}}
=
N(\mathcal{N})
\int_{T_{\mathrm{nr}}^{i}}^{T_{\mathrm{nr}}^{i+1}}
\hspace{-0.3cm}
d T_{\mathrm{nr}}
A(T_{\mathrm{nr}})
\int_{0}^{T^{\prime\text{max}}_{\text{nr}}}
\hspace{-0.3cm}
dT'_{\text{nr}}
\,
R(T_{\text{nr}},T'_{\text{nr}})
\int_{E_{\text{min}}(T'_{\text{nr}})}^{E_{\text{max}}}
\hspace{-0.3cm}
d E
\sum_{\nu=\nu_{e}, \nu_{\mu}, \bar{\nu}_{\mu}}
\frac{d N_{\nu}}{d E}(E)
\frac{d \sigma_{\nu-\mathcal{N}}}{d T_{\mathrm{nr}}}(E, T'_{\mathrm{nr}})
,
\end{equation}
where
$T_{\text{nr}}$ is the reconstructed nuclear recoil kinetic energy,
$T'_{\text{nr}}$ is the true nuclear recoil kinetic energy,
$A(T_{\text{nr}})$ is the energy-dependent detector efficiency,
$R(T_{\text{nr}},T'_{\text{nr}})$ is the energy resolution function,
$T^{\prime\text{max}}_{\text{nr}} = 2 E_{\text{max}}^2 / M$,
$E_{\text{max}} = m_\mu/2 \sim 52.8$ MeV,
$E_{\text{min}}(T'_{\text{nr}}) = \sqrt{MT'_\text{nr}/2}$,
$m_\mu$ being the muon mass, $M$ the nuclear mass,
and
$N(\mathcal{N})$ the number of $\mathcal{N}$ atoms in the detector.
We obtained information on these quantities from Refs.~\cite{COHERENT:2020iec,COHERENT:2020ybo} for the Ar data
and from Ref.~\cite{Akimov:2021dab} for the CsI data.
The number of $\mathcal{N}$ atoms in each detector is given by
$N(\mathcal{N}) = N_{\mathrm{A}} M_{\mathrm{det}} / M_{\mathrm{\mathcal{N}}}$, 
where $N_{\mathrm{A}}$ is the Avogadro number, 
$M_{\mathrm{det}}$ is the detector active mass
($M_{\mathrm{det}}=24 ~\mathrm{kg}$ for Ar
and
$M_{\mathrm{det}}=14.6 ~\mathrm{kg}$ for CsI),
and
$M_{\mathrm{\mathcal{N}}}$ is the molar mass
($M_{\mathrm{Ar}} = 39.96 ~\mathrm{g/mol}$
and
$M_{\mathrm{CsI}} = 259.8 ~\mathrm{g/mol}$).
The differential \cenns cross section
$d \sigma_{\nu-\mathcal{N}} / d T_{\mathrm{nr}}$
is discussed in Section~\ref{sec:cs}.

Due to the quenching effect, the energy actually observed is the electron-equivalent recoil energy $T_{e e}$, which is transformed into the nuclear recoil energy $T_{\mathrm{nr}}$  by inverting the relation
\begin{equation}\label{Qf}
	T_{e e}=f_{Q}\left(T_{\mathrm{nr}}\right) T_{\mathrm{nr}},
\end{equation}
where $f_{Q}$ is the quenching factor,
which is given in Refs.~\cite{Akimov:2021dab,COHERENT:2021pcd} for the CsI detector
and in Ref.~\cite{COHERENT:2020ybo} for the Ar detector.

An important characteristic of the neutrino beam in the COHERENT experiment
is the time dependence of the neutrino flavor components:
the prompt $\nu_{\mu}$'s produced in fast pion decay
($\tau_{\pi^{\pm}}\simeq26~\text{ns}$)
arrive within about $1~\mu\text{s}$ from the on-beam trigger,
whereas the delayed $\nu_{e}$'s and $\bar\nu_{\mu}$'s
produced in the slower muon decay
($\tau_{\mu^{\pm}}\simeq2.2~\mu\text{s}$)
arrive in a time interval which tails out at about $10~\mu\text{s}$.
Therefore, taking into account the time evolution of the data
is useful for distinguishing the interactions
of the two neutrino flavors.
We implemented the analyses of the COHERENT CsI and Ar data
using the timing information provided by the COHERENT
Collaboration~\cite{COHERENT:2020ybo,Akimov:2021dab,COHERENT:2021pcd}
and distributing the theoretical \cenns event numbers
$N^\text{\cenns}_{i}$ in Eq.~\eqref{N_cevns}
in time bins that are calculated from the exponential decay laws of
the generating pions and muons.
With this procedure we obtained the theoretical \cenns event numbers
$N^\text{\cenns}_{ij}$,
where $i$ is the index of the energy bins and $j$ is the index of the time bins.

We performed the analysis of the COHERENT CsI data in the energy and time bins
considered in Ref.~\cite{Akimov:2021dab}.
Since in some energy-time bins the number of events is zero,
we used the Poissonian least-squares function~\cite{Baker:1983tu,ParticleDataGroup:2020ssz}
\begin{align}
	\chi^2_{\mathrm{CsI}}
	=
	\null & \null
	2
	\sum_{i=1}^{9}
	\sum_{j=1}^{11}
	\left[
	    \sum_{z=1}^{4}( 1 + \eta_{z} ) N_{ij}^{z} -
		N_{ij}^{\text{exp}}
		+ N_{ij}^{\text{exp}} \ln\left(\frac{N_{ij}^{\text{exp}}}{\sum_{z=1}^{4}( 1 + \eta_{z} ) N_{ij}^{z}}\right)
	\right]
	+ \sum_{z=1}^{4}
	\left(
	\dfrac{ \eta_{z} }{ \sigma_{z} }
	\right)^2
	,
	\label{chi2coherentCsI}
\end{align}
where
the indices $i$ and $j$ denote, respectively, the energy and time bins,
and the indices
$z=1,2,3,4$ stand for \cenns,
beam-related neutron (BRN),
neutrino-induced neutron (NIN), and
steady-state (SS) backgrounds, respectively.
In our notation,
$N_{ij}^{\text{exp}}$ is the experimental event number obtained from coincidence (C) data,
$N_{ij}^{\text{\cenns}}$ is the predicted number of \cenns events
that depends on the physics model under consideration,
$N_{ij}^{\text{BRN}}$ is the estimated BRN background,
$N_{ij}^{\text{NIN}}$ is the estimated NIN background,
and
$N_{ij}^{\text{SS}}$ is the SS background obtained from the anti-coincidence (AC) data.
We took into account the systematic uncertainties described in
Ref.~\cite{Akimov:2021dab}
with the nuisance parameters $\eta_{z}$
and
the corresponding uncertainties
$\sigma_{\text{\cenns}}=0.12$
(which is the systematic uncertainty of the signal rate considering the effects of the 10\%, 3.8\%, 4.1\%, and 3.4\% uncertainties of the neutrino flux,
quenching factor,
\cenns efficiency,
and neutron form factors, respectively),
$\sigma_{\text{BRN}}=0.25$,
$\sigma_{\text{NIN}}=0.35$,
and
$\sigma_{\text{SS}}=0.021$.

We performed the analysis of the COHERENT Ar data in the energy and time bins
given in the data release~\cite{COHERENT:2020ybo}
with the least-squares function
\begin{align}
	\chi^2_{\mathrm{Ar}}
	=
	\null & \null
	\sum_{i=1}^{12}
	\sum_{j=1}^{10}
	\left(
	\dfrac
	{
		N_{ij}^{\text{exp}}
		- \sum_{z=1}^{4}( 1 + \eta_{z} 
		+ \sum_{l}\eta^{\mathrm{sys}}_{zl,ij} ) N_{ij}^{z}
	}
	{ \sigma_{ij} }
	\right)^2
	+ \sum_{z=1}^{4}
	\left(
	\dfrac{ \eta_{z} }{ \sigma_{z} }
	\right)^2
	+ \sum_{z,l}\left(
	\epsilon_{zl}
	\right)^2
	,
	\label{chi2coherentAr}
\end{align}
where $i$ is the index of the energy bins and $j$ is the index of the time bins.
Here $z=1,2,3,4$ stands for the theoretical prediction of \cenns, Steady-State (SS), Prompt Beam-Related Neutron (PBRN) and Delayed Beam-Related Neutron (DBRN) backgrounds, and $N_{ij}^{\text{exp}}$ is the number of observed events in each energy and time bin.
The statistical uncertainty ${\sigma_{ij}}$ is given by
\begin{equation}\label{sigma_stat}
	(\sigma_{ij}^\mathrm{} )^2 =
	( \sigma_{ij}^\mathrm{exp} )^2
	+ ( \sigma_{ij}^\text{SS} )^2,
\end{equation}
where 
$\sigma_{ij}^\mathrm{exp} = \sqrt{N_{ij}^{\text{exp}}}$ and 
$\sigma_{ij}^\mathrm{SS} = \sqrt{ {N_{ij}^{\text{SS}}}/{5} }$.
The factor 1/5 is due to the 5 times longer sampling time of the SS background with respect to the signal time window.   
The nuisance parameters $\eta_z$ quantify the systematic uncertainties of the event rate for the theoretical prediction of \cenns, SS, PBRN, and DBRN backgrounds,
with the corresponding uncertainties
$\sigma_\mathrm{CE\nu NS}=0.13$, 
$\sigma_\mathrm{PBRN}=0.32$, 
$\sigma_\mathrm{DBRN}=1$, and
$\sigma_\mathrm{SS}=0.0079$.
We considered also the systematic uncertainties of the shapes of \cenns and PBRN spectra
using the information in the COHERENT data release~\cite{COHERENT:2020ybo}.
This is done in Eq.~\eqref{chi2coherentAr} through
the nuisance parameters $\epsilon_{zl}$ and the terms $\eta^{\mathrm{sys}}_{zl,ij}$ given by
\begin{equation}\label{simga_sys}
	\eta^{\mathrm{sys}}_{zl,ij} = \epsilon_{zl} \, \frac{N_{zl,ij}^\mathrm{sys}-N_{zl,ij}^\mathrm{CV}}
	{N_{zl,ij}^\mathrm{CV}},
\end{equation}
where $l$ is the index of the source of the systematic uncertainty.
Here $N_{zl,ij}^\mathrm{sys}$ and $N_{zl,ij}^\mathrm{CV}$ are, respectively,
$1\sigma$ probability distribution functions (PDFs) described in Table~3 of Ref.~\cite{COHERENT:2020ybo}
and the central-value (CV) SM predictions described in Table~2 of Ref.~\cite{COHERENT:2020ybo}.
For the theoretical prediction of \cenns ($z=1$),
the sources of systematic shape uncertainties are the
$F_{90}$ energy dependence and the mean time to trigger ($t_\mathrm{trig}$) distribution.
For the PBRN background ($z=2$),
the sources of systematic shape uncertainties are the
energy, $t_\mathrm{trig}$ mean, and $t_\mathrm{trig}$ width distributions.

\section{CE\texorpdfstring{$\nu$}{}NS process and light mediators}
\label{sec:cs}

In the SM,
the differential cross section as a function of the nuclear kinetic recoil energy $T_{\text{nr}}$ of the \cenns process with a neutrino $\nu_{\ell}$ ($\ell=e,\mu,\tau$) and a nucleus $\mathcal{N}$ is given by~\cite{Drukier:1984vhf,Barranco:2005yy,Patton:2012jr}
\begin{equation}
	\dfrac{d\sigma_{\nu_{\ell}\text{-}\mathcal{N}}}{d T_{\text{nr}}}
	(E,T_{\text{nr}})
	=
	\dfrac{G_{\text{F}}^2 M}{\pi}
	\left( 1 - \dfrac{M T_{\text{nr}}}{2 E^2} \right)
	(Q^{V}_{\ell, \mathrm{SM}})^2,
	\label{cs-std}
\end{equation}
where $G_{\text{F}}$ is the Fermi constant and
\begin{equation}\label{Qsm}
	Q^{V}_{\ell, \mathrm{SM}}=\left[g_{V}^{p}\left(\nu_{\ell}\right) Z F_{Z}(|\vec{q}|^{2})+g_{V}^{n} N F_{N}(|\vec{q}|^{2})\right]^{},
\end{equation}
is the weak charge of the nucleus.
Here, $Z$ and $N$ are the numbers of protons and neutrons in the nucleus, respectively,
and $g_{V}^{p}$ and $g_{V}^{n}$ are the neutrino-proton and neutrino-neutron couplings, respectively.
Taking into account radiative corrections in the $\overline{\mathrm{MS}}$ scheme~\cite{Erler:2013xha}, accurate values of the vector couplings can be derived as~\cite{Cadeddu:2020lky}\footnote{
A different treatment of the hadronic uncertainties is discussed in Refs.~\cite{Tomalak:2020zfh,Crivellin:2021bkd}.
The resulting small differences for the values of
$g_{V}^{p}$ and $g_{V}^{n}$
can be neglected in the current analyses of \cenns data
which have other large uncertainties.
}
\begin{align}
	g_{V}^{p}(\nu_{e}) = 0.0401,\quad\quad
	g_{V}^{p}(\nu_{\mu}) = 0.0318,\quad\quad
	g_{V}^{n} = -0.5094\,.
	\label{gVn}
\end{align}
In Eq.~\eqref{Qsm},
$F_{Z}(|\vec{q}|^{2})$ and $F_{N}(|\vec{q}|^{2})$ are, respectively, the form factors of the proton and neutron distributions in the nucleus, which are the Fourier transforms of the corresponding nucleon distribution in the nucleus and describe the loss of coherence for large values of the momentum transfer $|\vec{q}|$.
We use an analytic expression, namely the Helm parameterization~\cite{Helm:1956zz}, for the form factors, that gives practically equivalent results to the other two well known parameterizations, i.e., the symmetrized Fermi~\cite{Piekarewicz:2016vbn} and Klein-Nystrand~\cite{Klein:1999qj} ones.
The proton rms radii can be obtained from the muonic atom spectroscopy experiments~\cite{Fricke:1995zz,Angeli:2013epw} as explained in Ref.~\cite{Cadeddu:2020lky}
\begin{align}\label{Rp}
	R_{p}(\mathrm{Cs})=4.821~\mathrm{fm}, \quad\quad R_{p}(\mathrm{I})=4.766 ~\mathrm{fm},
	\quad\quad
	R_{p}(\mathrm{Ar})=3.448~\mathrm{fm}.
\end{align}
On the other hand,
there is a poor knowledge of the values of the ${}^{133}\text{Cs}$, ${}^{127}\text{I}$ and ${}^{40}\text{Ar}$
neutron rms radii obtained from the analyses of the COHERENT data~\cite{Cadeddu:2017etk,Papoulias:2019lfi,Cadeddu:2018dux,Huang:2019ene,Papoulias:2019txv,Khan:2019cvi,Cadeddu:2020lky,Cadeddu:2019eta}.
The values of these neutron rms radii can, however, be estimated with theoretical calculations
based on different nuclear models~\cite{Hoferichter:2020osn,Cadeddu:2020lky,Cadeddu:2021ijh}.
Here, we consider the following values obtained from the recent nuclear shell model
estimate of the corresponding neutron skins
(i.e. the differences between the neutron and the proton rms radii)
in Ref.~\cite{Hoferichter:2020osn}
\begin{align}\label{Rn}
	R_{n}(\mathrm{Cs})\simeq5.09~\mathrm{fm}, \quad\quad R_{n}(\mathrm{I})\simeq5.03~\mathrm{fm},\quad\quad
	R_{n}(\mathrm{Ar})\simeq3.55~\mathrm{fm}.
\end{align}
Following the COHERENT Collaboration~\cite{Akimov:2021dab,COHERENT:2020iec,COHERENT:2020ybo},
we take into account the effect of the uncertainty of the values of the neutron rms radii by considering
3.4\% and 2\% uncertainties for the CsI and Ar \cenns rates, respectively.

The SM \cenns differential event rates that are predicted for the COHERENT Ar and CsI detectors
are shown in Fig.~\ref{fig:spe_vector} as functions of $T_{\text{nr}}$.
One can see that there are kinks at
$T_{\text{nr}} \approx 50~\text{keV}$ for Ar
and
$T_{\text{nr}} \approx 15~\text{keV}$ for CsI.
The steeper slope of the SM differential event rates below these values of
$T_{\text{nr}}$
is due to the coherency condition
$T_{\text{nr}} \lesssim 1 / 2 M R^2$.


The \cenns cross section is modified if there is a new massive mediator which couples to the SM leptons and quarks. In this work,
we focus on two mediator types that have been considered in several previous works~\cite{Liao:2017uzy,Papoulias:2017qdn,Papoulias:2018uzy,Papoulias:2019xaw,Khan:2019cvi,Dutta:2019eml,Bertuzzo:2017tuf,Abdullah:2018ykz,Billard:2018jnl,Han:2019zkz,Flores:2020lji,Papoulias:2019txv,Miranda:2020tif,Cadeddu:2020nbr,delaVega:2021wpx,Bertuzzo:2021opb,CONUS:2021dwh,Bauer:2018onh,Bauer:2020itv,Amaral:2020tga}:
an additional vector mediator $Z'$ with mass $M_{Z^{\prime}}$ associated to a new $U(1)'$ gauge group
and
an additional scalar mediator $\phi$ with mass $M_{\phi}$.
The phenomenology of \cenns in the specific models that we consider is briefly described in the following two Subsections.

\subsection{Light vector mediator}
\label{subs:vector_mediator}

The interaction of a $Z'$ vector boson with neutrinos and quarks is described by the generic Lagrangian
\begin{equation}
\mathcal{L}_{Z'}^{V}
=
-
Z'_{\mu}
\left[
\sum_{\ell=e,\mu,\tau}
g_{Z'}^{\nu_{\ell}V}
\,
\overline{\nu_{\ell L}} \gamma^{\mu} \nu_{\ell L}
+
\sum_{q=u,d}
g_{Z'}^{q V}
\,
\overline{q} \gamma^{\mu} q
\right]
,
\label{LZp}
\end{equation}
where $g_{Z^{\prime}}^{q V}$ and $g_{Z^{\prime}}^{\nu_{\ell}V}$
are the couplings constants.

In the case of a vector mediator associated with a new $U(1)'$ gauge group,
the coupling constants are proportional to the charges
$Q^{\prime}_{q}$ and $Q^{\prime}_{\ell}$
of quarks and neutrinos under the new gauge symmetry:
$g_{Z^{\prime}}^{q V} = g_{Z^{\prime}} Q^{\prime}_{q}$
and
$g_{Z^{\prime}}^{\nu_{\ell}V} = g_{Z^{\prime}} Q^{\prime}_{\ell}$,
where $g_{Z^{\prime}}$ is the coupling constant of the symmetry group.
Since both the SM and the $Z'$ interactions are of vector type,
they contribute coherently to the \cenns cross section.
Moreover,
since the vector current is conserved, the proton and neutron coupling are given
by the sums of the couplings of their valence quarks.
Therefore, the total cross section is obtained by replacing the SM weak charge $Q^{V}_{\ell,\mathrm{SM}}$
with the new total weak charge (see Appendix~\ref{app:coupling})
\begin{equation}\label{Q_ll}
	Q^{V}_{\ell,\mathrm{SM+V}}=
	Q^{V}_{\ell,\mathrm{SM}} + 
	\frac{ g^2_{Z^{\prime}} Q_{\ell}^{\prime} }{ \sqrt{2} G_{F}\left(|\vec{q}|^{2}+M_{Z^{\prime}}^{2}\right)} 
	\left[ 
	 \left(2 Q_{u}^{\prime} +  Q_{d}^{\prime}\right) Z F_{Z}(|\vec{q}|^{2})
	 +  \left( Q_{u}^{\prime} +  2Q_{d}^{\prime}\right) N F_{N}(|\vec{q}|^{2})
	\right]
	,
\end{equation}
with
$|\vec{q}|^2 \simeq 2 M T_{\text{nr}}$.

In this work we consider the models listed in Table~\ref{tab:charges}.
There are many models beyond the SM with an additional massive $Z'$ vector boson associated with a new $U(1)'$ gauge symmetry (see, e.g., the review in Ref.~\cite{Langacker:2008yv}).
A necessary requirement is that the theory is anomaly-free.
However, it is possible to consider effective anomalous models that describe the interactions of SM fermions with the implicit requirement that the anomalies are canceled by the contributions of the non-standard fermions of the full theory.
This is the case of the first model that we consider: a $Z'$ boson which couples universally to all SM fermions~\cite{Liao:2017uzy,Papoulias:2017qdn,Billard:2018jnl,Papoulias:2019txv,Khan:2019cvi,Cadeddu:2020nbr,Bertuzzo:2021opb}.
In this case $Q^{\prime}_{\ell}=Q^{\prime}_{u}=Q^{\prime}_{d}=1$, and the coupling
is same for all the fermions.

\begin{table}[!t]
\renewcommand{\arraystretch}{1.45}
    \centering
    \begin{tabular}{c|cc|rrr}
    \hline\hline 
    Model & $Q'_{u}$ & $Q'_{d}$  & $Q'_{e}$ & $Q'_{\mu}$ & $Q'_{\tau}$ \\
    \hline
    universal & 1 & 1  & 1 & 1 & 1 \\
    \hline
    $B-L$ & $1/3$ & $1/3$  & $-1$ & $-1$ & $-1$ \\
    $B-3 L_{e}$ & $1/3$ & $1/3$  & $-3$ & 0 & 0 \\
    $B-3 L_{\mu}$ & $1/3$ & $1/3$  & 0 & $-3$ & 0 \\
    $B-2L_{e}-L_{\mu}$ & $1/3$ & $1/3$ & $-2$ & $-1$ & 0 \\
    $B-L_{e}-2L_{\mu}$ & $1/3$ & $1/3$ & $-1$ & $-2$ & 0 \\
    $B_y+L_\mu+L_\tau$ 
    & $1/3$ & $1/3$  & 0 & 1 & 1  \\
    \hline
    $L_{e}-L_{\mu}$    & 0 & 0 & 1 & $-1$ & 0  \\
    $L_{e}-L_{\tau}$   & 0 & 0 & 1 & 0  & $-1$  \\
    $L_{\mu}-L_{\tau}$ & 0 & 0 & 0 & 1  & $-1$  \\
    \hline\hline
    \end{tabular}
    \caption{The $U(1)'$ charges of quarks and leptons in the vector mediator models considered in this work.}
    \label{tab:charges}
\end{table}

Other models that we consider are anomaly-free if the SM is extended with the introduction of three right-handed neutrinos
(see, e.g., Ref.~\cite{Allanach:2018vjg}), which are also beneficial for the generation of the neutrino masses
that are necessary for the explanation of the oscillations of neutrinos observed in many experiments (see, e.g., Refs.~\cite{Giunti:2007ry,ParticleDataGroup:2020ssz}).
In this case,
there is an infinite set of anomaly-free $U(1)'$ gauge groups generated by
\begin{equation}
G(c_{1},c_{2},c_{3},c_{e},c_{\mu},c_{\tau})
=
c_{1} B_{1} + c_{2} B_{2} + c_{3} B_{3} - c_{e} L_{e} - c_{\mu} L_{\mu} - c_{\tau} L_{\tau}
,
\label{G1}
\end{equation}
where
$B_{1}$, $B_{2}$, and $B_{3}$ are the baryon numbers of the three generations and $L_{\alpha}$ are the lepton numbers for $\alpha=e,\mu,\tau$.
We assume that for each generation the $U(1)'$ couplings of the right-handed neutrino is the same as that of the left-handed neutrino in order to have
vectorial $U(1)'$ interactions.
Therefore,
when we extend the SM gauge group
$SU(3)_{C} \times SU(2)_{L} \times U(1)_{Y}$
to
$SU(3)_{C} \times SU(2)_{L} \times U(1)_{Y} \times U(1)'$,
there are no
$[SU(3)_{C}]^2 U(1)_{Y}$,
$[U(1)']^3$
and
$[\text{gravity}]^2 U(1)'$
anomalies,
because of the vectorial character of the involved interactions.
The $[U(1)']^2 U(1)_{Y}$
anomaly cancels because
for each generation the difference of the $Y$ charges of
left-handed and right-handed quarks (leptons) is zero.
The remaining
$[SU(2)_{L}]^2 U(1)'$
and
$[U(1)_{Y}]^2 U(1)'$
anomalies are canceled with the constraint
\begin{equation}
c_{1} + c_{2} + c_{3} - c_{e} - c_{\mu} - c_{\tau} = 0
.
\label{c1}
\end{equation}

It is often assumed that the quark charges are universal,
in order to avoid unobserved flavor-changing neutral currents in the quark sector.
In this case,
we have
\begin{equation}
G_{B}(c_{B},c_{e},c_{\mu},c_{\tau})
=
c_{B} B - c_{e} L_{e} - c_{\mu} L_{\mu} - c_{\tau} L_{\tau}
,
\label{G2}
\end{equation}
with the constraint (see, e.g., Refs.~\cite{Lee:2010hf,Araki:2012ip})
\begin{equation}
3 c_{B} - c_{e} - c_{\mu} - c_{\tau} = 0
.
\label{c2}
\end{equation}
Here $ B = B_{1} + B_{2} + B_{3} $ is the usual baryon number.

We consider the following anomaly-free models that correspond to different choices of the coefficients in Eq.~\eqref{G1} or~\eqref{G2}
and contribute to \cenns interactions of $\nu_e$ and $\nu_\mu$:

\begin{description}

\item[$\bm{B-L=G_{B}(1,1,1,1)}$]
Here $ L = L_{e} + L_{\mu} + L_{\tau} $ is the total lepton number.
This is the most popular $Z'$ model, with a huge literature
(see, e.g., the reviews in Refs.~\cite{Langacker:2008yv,Mohapatra:2014yla,Okada:2018ktp}).
It was considered recently in several \cenns phenomenological analyses,
e.g. those in Refs.~\cite{Miranda:2020tif,Cadeddu:2020nbr,Coloma:2020gfv,delaVega:2021wpx,Bertuzzo:2021opb}.
Note that,
since there are no $\nu_\tau$'s in the COHERENT neutrino beam,
bounds on the coupling constant in the anomaly-free model
generated by
\begin{equation}
G_{B}(1,3/2,3/2,0) = B - \frac{3}{2} \left( L_e + L_\mu \right)
,
\label{Bm32}
\end{equation}
considered, e.g., in Ref.~\cite{Han:2019zkz},
can be obtained from the bounds on the coupling constant $g_{Z'}$
in the $B-L$ model by rescaling it by the factor $\sqrt{2/3}$,
because the $\nu_e$ and $\nu_\mu$ couplings are changed by the same factor $3/2$.

\item[$\bm{B_y+L_\mu+L_\tau=G(1,-y,y-3,0,-1,-1)}$]
In this model,
proposed in Ref.~\cite{Farzan:2015doa}
and considered, e.g., in Ref.~\cite{Coloma:2020gfv},
$ B_y = B_1 - y B_2 + ( y - 3 ) B_3 $.

\item[$\bm{B-3L_e=G_{B}(1,3,0,0)}$]
This model was considered, e.g., in Refs.~\cite{Han:2019zkz,Heeck:2018nzc,Coloma:2020gfv,delaVega:2021wpx}.
In this case,
only the $\nu_e$ \cenns cross section is affected by the new $Z'$-mediated interaction.
Moreover, since there are no $\nu_\tau$'s in the COHERENT neutrino beam,
the bounds on the coupling constant $g_{Z'}$
obtained in this model can be extended to all the anomaly-free models generated by
\begin{equation}
G_{B}(1,3w_e,0,3(1-w_e)) = B - 3 w_e L_e - 3 (1-w_e) L_\tau
\label{Bm3wLe}
\end{equation}
through a rescaling of the coupling constant by a factor $1/\sqrt{w_e}$.

\item[$\bm{B-3L_\mu=G_{B}(1,0,3,0)}$]
This model was considered, e.g., in Refs.~\cite{Heeck:2018nzc,Coloma:2020gfv,delaVega:2021wpx}.
In this case,
only the $\nu_\mu$ \cenns cross section is affected by the new $Z'$-mediated interaction
and,
in analogy with the argument in the previous item,
the bounds on the coupling constant $g_{Z'}$
obtained in this model can be extended to all the anomaly-free models generated by
\begin{equation}
G_{B}(1,0,3w_\mu,3(1-w_\mu)) = B - 3 w_\mu L_\mu - 3 (1-w_\mu) L_\tau
\label{Bm3wLmu}
\end{equation}
through a rescaling of the coupling constant by a factor $1/\sqrt{w_\mu}$.
For example, the $B-(3/2)(L_\mu+L_\tau)$ considered in Refs.~\cite{Heeck:2018nzc,Coloma:2020gfv}
is obtained with $w_\mu=1/2$.

\item[$\bm{B-2L_e-L_\mu=G_{B}(1,2,1,0)}$]
This model was considered, e.g., in Ref.~\cite{delaVega:2021wpx}.
In analogy with the discussion in the previous items,
the bounds on the coupling constant $g_{Z'}$
obtained in this model can be extended to all the anomaly-free models generated by
\begin{equation}
G_{B}(1,2w_1,w_1,3(1-w_1)) = B - 2 w_1 L_e - w_1 L_\mu - 3 (1-w_1) L_\tau
\label{Bm3wL1}
\end{equation}
through a rescaling of the coupling constant by a factor $1/\sqrt{w_1}$.

\item[$\bm{B-L_e-2L_\mu=G_{B}(1,1,2,0)}$]
This model, was considered, e.g., in Ref.~\cite{delaVega:2021wpx}.
Again, in analogy with the discussion in the previous items,
the bounds on the coupling constant $g_{Z'}$
obtained in this model can be extended to all the anomaly-free models generated by
\begin{equation}
G_{B}(1,w_2,2w_2,3(1-w_2)) = B - w_2 L_e - 2 w_2 L_\mu - 3 (1-w_2) L_\tau
\label{Bm3wL2}
\end{equation}
through a rescaling of the coupling constant by a factor $1/\sqrt{w_2}$.

\end{description}

\begin{figure}[!t]
\centering
\subfigure[]{\label{fig:spe_vector_a}
\includegraphics[width=0.48\textwidth]{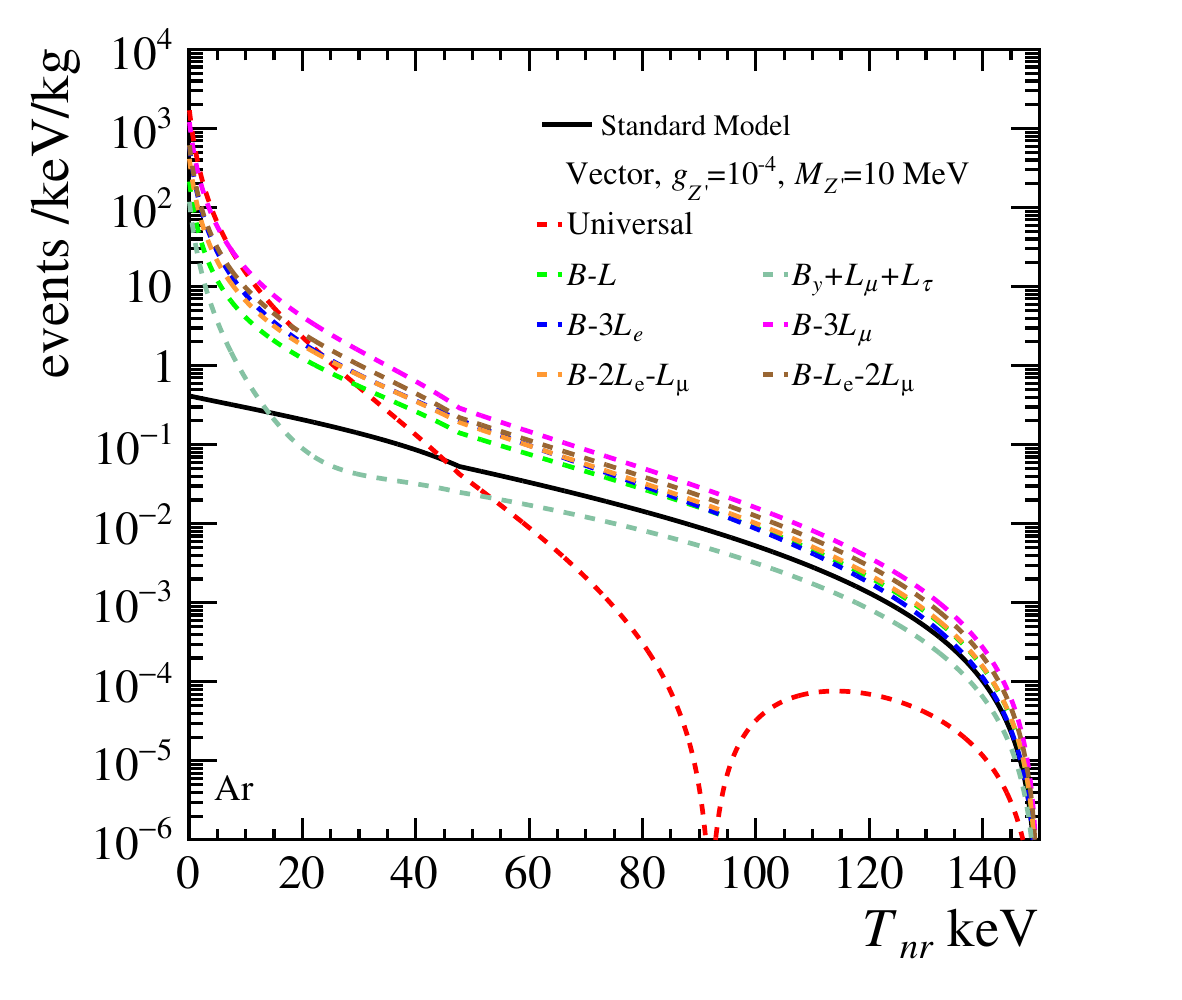}
}
\subfigure[]{\label{fig:spe_vector_b}
\includegraphics[width=0.48\textwidth]{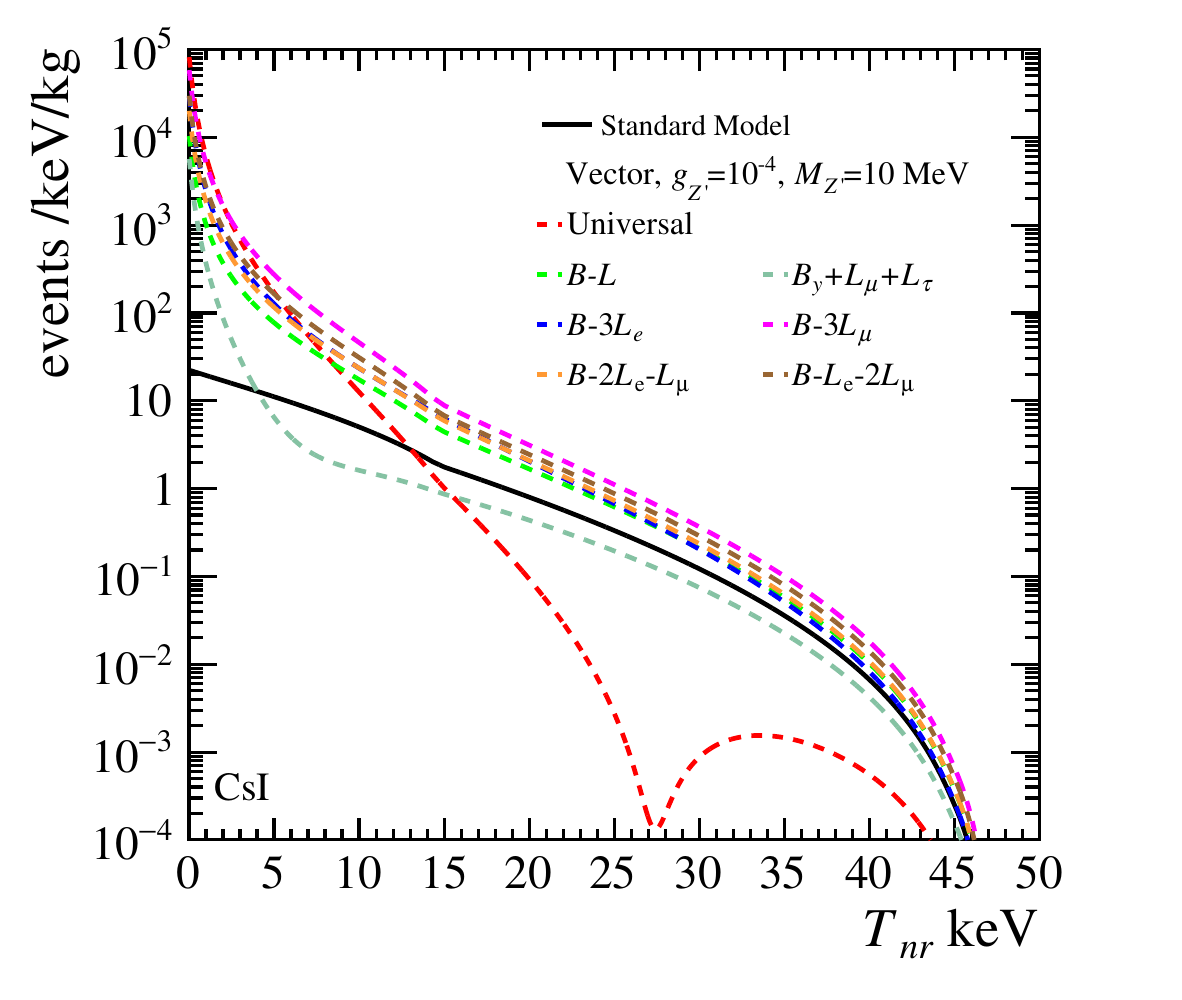}
}
\vspace{-0.5 cm}
\subfigure[]{\label{fig:spe_vector_c}
\includegraphics[width=0.48\textwidth]{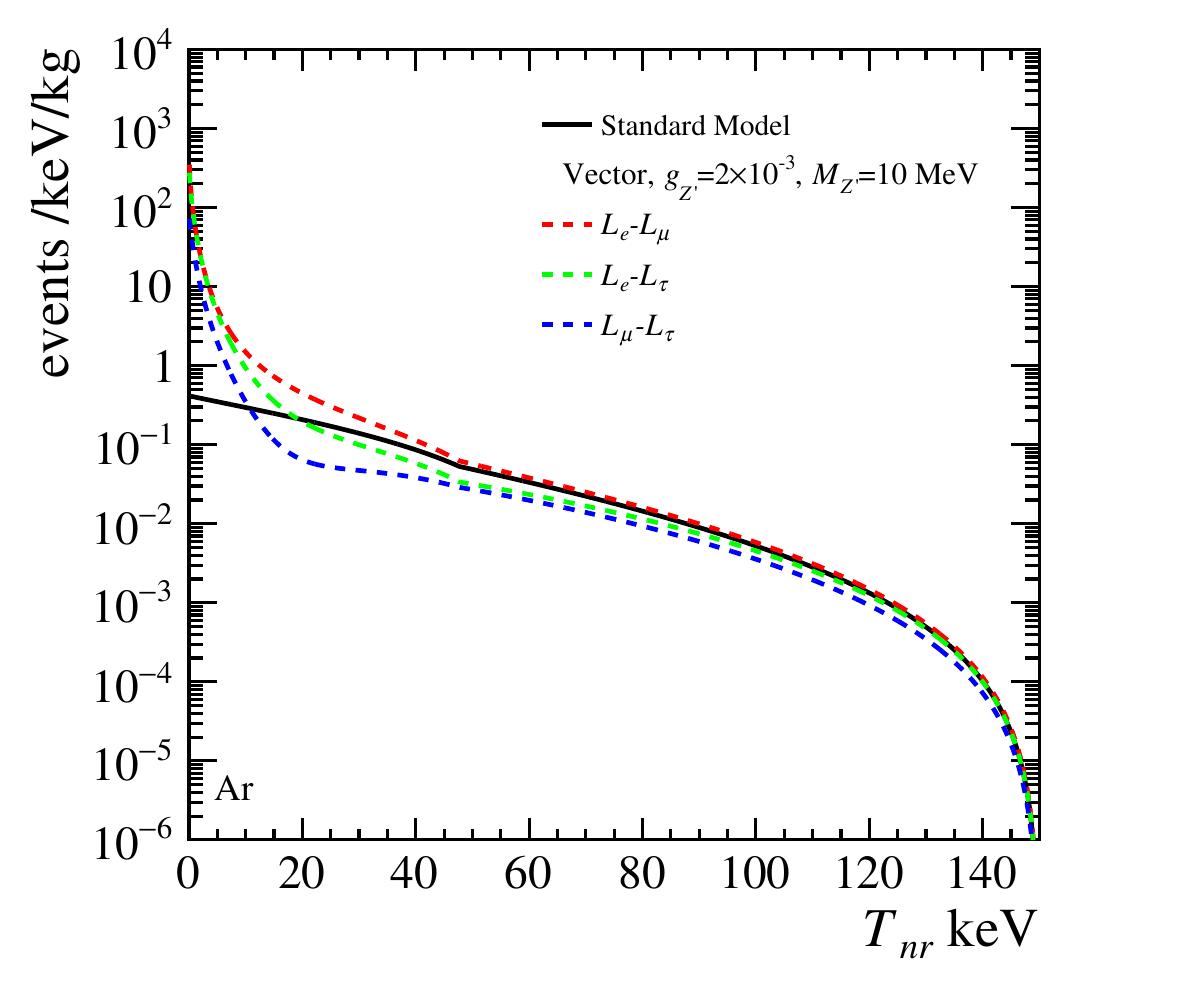}
}
\subfigure[]{\label{fig:spe_vector_d}
\includegraphics[width=0.48\textwidth]{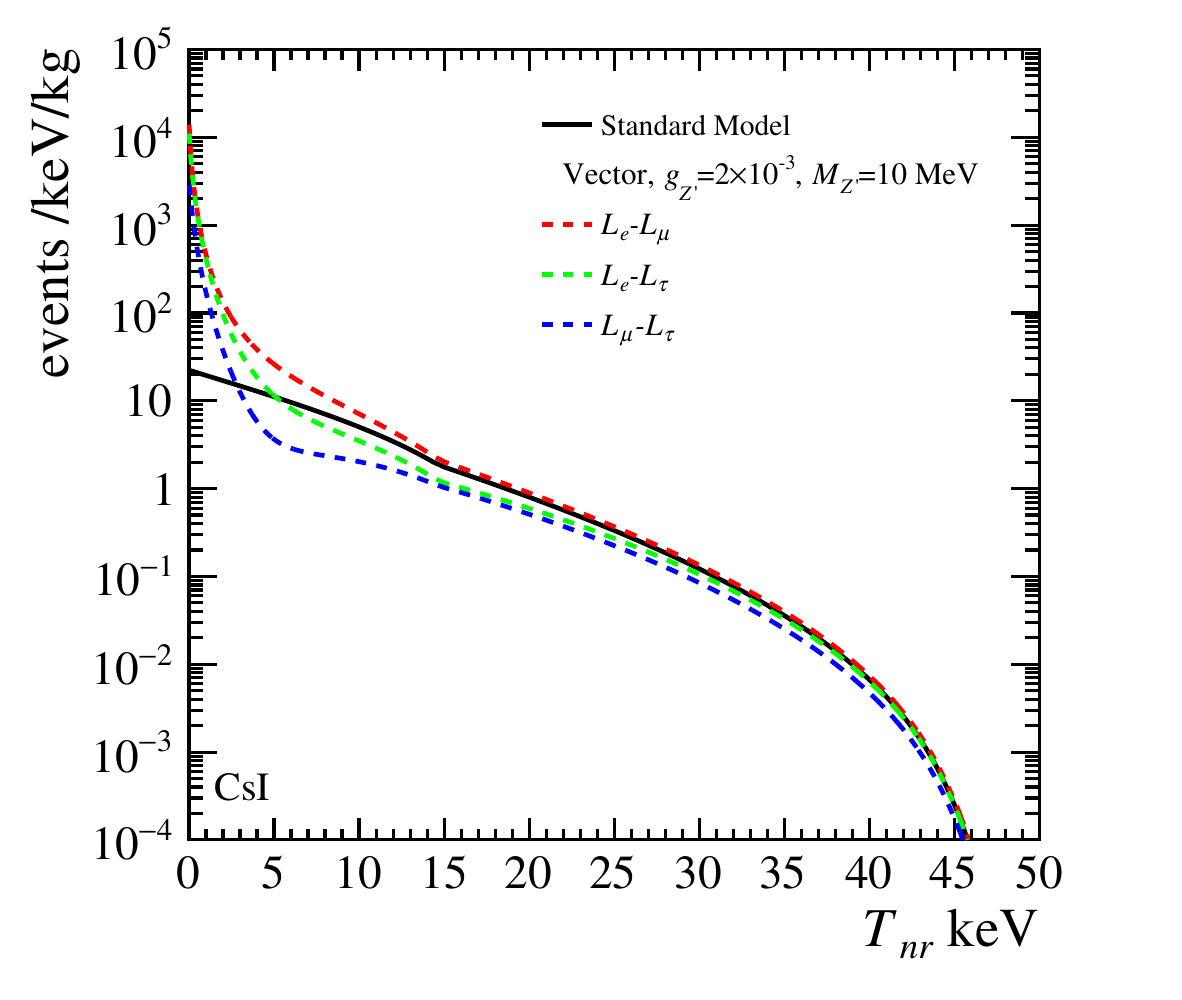}
}
\caption{Predicted CE$\nu$NS differential event rates corresponding to the experimental configuration and data taking time of the COHERENT Ar (a, c) and CsI (b, d) detectors in the vector mediator models considered in this work.}
\label{fig:spe_vector}
\end{figure}

The effects of the models above on the \cenns differential event rates that are predicted for the COHERENT Ar and CsI detectors
are illustrated, respectively, in Figs.~\ref{fig:spe_vector_a} and~\ref{fig:spe_vector_b}.
In these figures we choose $g_{Z'}=10^{-4}$ and $M_{Z'}=10~\text{MeV}$
and we compared the model predictions with the SM one.
One can see that
the effects of the light mediator are similar for the Ar and CsI detectors
and the vector boson mediator contribution increases for small values of $T_{\text{nr}} \simeq |\vec{q}|^2 / 2 M$
because of the propagator in Eq.~\eqref{Q_ll}.
The different scales of $T_{\text{nr}}$ in Figs.~\ref{fig:spe_vector_a} and~\ref{fig:spe_vector_b} are obviously due to the different masses of the nuclei.

In the case of the universal $Z'$ model there is a deep dip due to a cancellation between the negative SM and the positive $Z'$ contributions to the weak charge in Eq.~\eqref{Q_ll}.
This occurs only in the universal model because only in this case
all the quark and lepton charges are positive and both $\nu_e$ and $\nu_\mu$
interact with the $Z'$.
Indeed,
there is a cancellation for
\begin{equation}
T_{\text{nr}}
=
-
\frac{1}{2M}
\left(
\frac{ 3 g_{Z'}^2 }{  \sqrt{2} G_{F} }
\,
\frac{ Z F_{Z}(|\vec{q}|^{2})
     + N F_{N}(|\vec{q}|^{2}) }
     { g_{V}^{p} Z F_{Z}(|\vec{q}|^{2})
     + g_{V}^{n} N F_{N}(|\vec{q}|^{2}) }
+
M_{Z'}^2
\right)
,
\label{univ_canc}
\end{equation}
which occurs at
$ T_{\text{nr}} \simeq 92~\text{keV} $ for Ar in Fig.~\ref{fig:spe_vector_a}
and
$ T_{\text{nr}} \simeq 27~\text{keV} $ for CsI in Fig~\ref{fig:spe_vector_b}.

There is a cancellation for $\nu_\mu$ also in the $B_y+L_\mu+L_\tau$ model,
since the quarks and $\nu_{\mu}$ have positive charges
(see Table~\ref{tab:charges}).
The cancellation occurs at
\begin{equation}
T_{\text{nr}}
=
-
\frac{1}{2M}
\left(
\frac{ g_{Z'}^2 }{ \sqrt{2} G_{F} }
\,
\frac{ Z F_{Z}(|\vec{q}|^{2})
     + N F_{N}(|\vec{q}|^{2}) }
     { g_{V}^{p} Z F_{Z}(|\vec{q}|^{2})
     + g_{V}^{n} N F_{N}(|\vec{q}|^{2}) }
+
M_{Z'}^2
\right)
,
\label{bylm_canc}
\end{equation}
which corresponds to
$ T_{\text{nr}} \simeq 29~\text{keV} $ for Ar in Fig.~\ref{fig:spe_vector_a}
and
$ T_{\text{nr}} \simeq 8~\text{keV} $ for CsI in Fig.~\ref{fig:spe_vector_b}.
Since in this case there is no cancellation of the SM contribution of $\nu_e$,
which does not interact with the $Z'$,
there are only shallow dips at these energies in Figs.~\ref{fig:spe_vector_a} and~\ref{fig:spe_vector_b} for this model.
Note that the total differential rate is smaller than the SM differential rate for energies above the dip,
because the positive and smaller $Z'$ contribution to $Q^{V}_{\mu,\mathrm{SM+V}}$
is added to the dominant negative SM contribution,
decreasing the absolute value of $Q^{V}_{\mu,\mathrm{SM+V}}$.

In all the other models above the quarks and leptons have opposite charges
(see Table~\ref{tab:charges})
and the $Z'$ contribution to the weak charge in Eq.~\eqref{Q_ll}
is negative as the SM contribution.
Therefore,
the total differential rate is larger than the SM rate
for all values of $T_{\text{nr}}$,
as shown in 
Figs.~\ref{fig:spe_vector_a} and~\ref{fig:spe_vector_b}.

We also consider the following three possible
$L_\alpha-L_\beta$
models that are anomaly-free and can be gauged without extending the SM content with right-handed neutrinos~\cite{Foot:1990mn,Foot:1990uf,He:1990pn,Foot:1992ui}:

\begin{description}

\item[$\bm{L_e-L_\mu=G_{B}(0,-1,1,0)}$]
This model, obtained from Eq.~\eqref{G2} with
$c_{B} = 0$,
$c_{e} = -1$
$c_{\mu} = 1$, and
$c_{\tau} = 0$,
was considered, e.g., in Refs.~\cite{He:1990pn,He:1991qd,Coloma:2020gfv}.

\item[$\bm{L_e-L_\tau=G_{B}(0,-1,0,1)}$]
This model, obtained from Eq.~\eqref{G2} with
$c_{B} = 0$,
$c_{e} = -1$
$c_{\mu} = 0$, and
$c_{\tau} = 1$,
was considered, e.g., in Refs.~\cite{He:1990pn,He:1991qd,Coloma:2020gfv}.

\item[$\bm{L_\mu-L_\tau=G_{B}(0,0,-1,1)}$]
This model, obtained from Eq.~\eqref{G2} with
$c_{B} = 0$,
$c_{e} = 0$
$c_{\mu} = -1$, and
$c_{\tau} = 1$,
was considered in many papers, e.g., in Refs.~\cite{He:1990pn,Baek:2001kca,Altmannshofer:2019zhy,Banerjee:2018mnw,Gninenko:2020xys,Cadeddu:2020nbr,Banerjee:2021laz,Bertuzzo:2021opb}.

\end{description}

\begin{figure}[!t]
\begin{center}
\includegraphics*[width=0.48\linewidth]{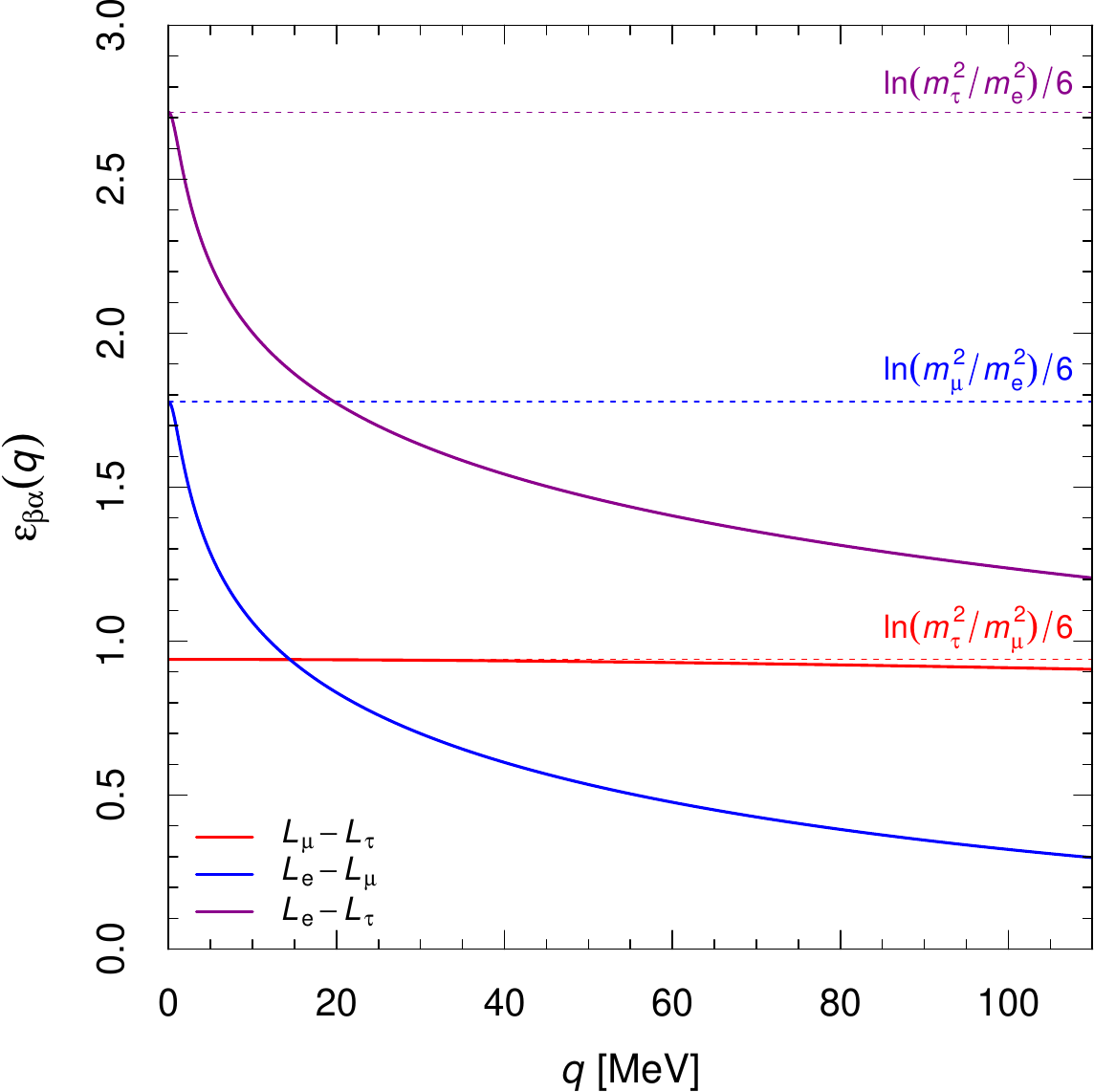}
\caption{ \label{fig:loopint}
Values of $\varepsilon_{\beta\alpha}$ in Eq.~\eqref{eps} for each of the three
$L_\alpha-L_\beta$
symmetries
as a function of $ q = |\vec{q}| \simeq \sqrt{2 M T_{\mathrm{nr}}} $
in the range of the COHERENT \cenns data.
}
\end{center}
\end{figure}

Since in these models the $Z'$ vector boson does not couple to quarks,
there are no tree-level interactions that contribute to \cenns
(assuming the absence of tree-level kinetic mixing).
However,
there is kinetic mixing of the $Z'$ and the photon at the one-loop level that induces a contribution to \cenns through the photon interaction with quarks
\cite{Altmannshofer:2019zhy,Banerjee:2018mnw,Banerjee:2021laz}.
The \cenns cross section in these three models is~\cite{Altmannshofer:2019zhy,Cadeddu:2020nbr}\footnote{
We correct here the sign of the $Z'$ contribution with respect to that used in Ref.~\cite{Cadeddu:2020nbr}.
Let us also note that in the analysis in Ref.~\cite{Banerjee:2021laz} the $Z'$ contribution has the correct sign, but there is an additional factor $1/2$ that is incorrect, as shown in Appendix~\ref{app:coupling}.
}
\begin{align}
  \left(\frac{d \sigma}{d T_{\mathrm{nr}}}\right)_{L_{\alpha}-L_{\beta}}^{\nu_{\ell}-\mathcal{N}}
  &
  \left(E, T_{\mathrm{nr}}\right)
=
\frac{G_{F}^{2} M}{\pi}\left(1-\frac{M T_{\mathrm{nr}}}{2 E^{2}}\right)
\nonumber
\\
  \times
&
  \left\{
  \left[
  g_{V}^{p}\left(\nu_{\ell}\right)
  +
  \frac{\sqrt{2}\alpha_{\mathrm{EM}} g_{Z^{\prime}}^{2}
        \left( \delta_{\ell\alpha} \varepsilon_{\beta\alpha}(|\vec{q}|)
             + \delta_{\ell\beta} \varepsilon_{\alpha\beta}(|\vec{q}|) \right)}
       {\pi G_{F} \left(|\vec{q}|^{2}+M_{Z^{\prime}}^2\right)}
  \right] Z F_{Z}(|\vec{q}|^{2})
  +
  g_{V}^{n} N F_{N}(|\vec{q}|^{2})
  \right\}^{2}
  ,
\label{cs_mu_tau}
\end{align}
where $\alpha_{\mathrm{EM}}$ is the electromagnetic fine-structure constant and $\varepsilon_{\beta\alpha}(|\vec{q}|)$ is the one-loop kinetic mixing coupling,
that is given by~\cite{Banerjee:2018mnw,Banerjee:2021laz}
\begin{equation}
    \varepsilon_{\beta\alpha}(|\vec{q}|) =
    \int_{0}^{1} x(1-x)
    \ln\left(
    \frac{ m_{\beta}^{2} + x(1-x) |\vec{q}|^{2} }
         {m_{\alpha}^{2} + x(1-x) |\vec{q}|^{2} }
    \right)
    d x\,,
\label{eps}
\end{equation}
where $m_{\beta}$ and $m_{\alpha}$ are the charged lepton masses
and we took into account that for \cenns
$q^2 \simeq - |\vec{q}|^2 \simeq - 2 M T_{\mathrm{nr}}$.
Note that the $Z'$ contribution is invariant for $\alpha \leftrightarrows \beta$,
as it should be, since $L_\alpha-L_\beta$ and $L_\beta-L_\alpha$
are physically equivalent.
Note also that
the sign of the loop contribution of the $i$ charged lepton
to $\nu_\ell$ scattering is given by
$- Q'_i Q'_\ell$,
where the minus comes from the negative electric charge of the charged lepton
propagating in the loop.
Therefore, the mass of the charged lepton with the same flavor $\ell$ of the
scattering neutrino is always at the denominator of the logarithm in Eq.~\eqref{eps}
and the mass of the other charged lepton taking part to the new symmetry
is always at the numerator.
Figure~\ref{fig:loopint}
shows the value of $\varepsilon_{\beta\alpha}(|\vec{q}|)$ for each of the three
$L_\alpha-L_\beta$
symmetries
as a function of $|\vec{q}|$
in the range of the COHERENT \cenns.
One can see that only $\varepsilon_{\tau\mu}$
is almost constant, because $|\vec{q}| \ll m_\tau$ and $|\vec{q}| < m_\mu$.
In this case it is possible to approximate
$\varepsilon_{\tau\mu}\simeq\ln(m_\tau^2/m_\mu^2)/6$,
as done in Refs.~\cite{Altmannshofer:2019zhy,Cadeddu:2020nbr,Bertuzzo:2021opb}.
On the other hand,
for the symmetries
$L_e-L_\mu$ and $L_e-L_\tau$ the $|\vec{q}|$ dependence of $\varepsilon_{\beta\alpha}$
on $|\vec{q}|$ must be taken into account, because $|\vec{q}| \gg m_{e}$.

Figures~\ref{fig:spe_vector_c} and~\ref{fig:spe_vector_d}
illustrate the effects of the $Z'$ contribution
to the \cenns differential event rates that are predicted for the COHERENT Ar and CsI detectors
in the $L_\alpha-L_\beta$ models.
In these figures we choose $g_{Z'}=2\times10^{-3}$ and $M_{Z'}=10~\text{MeV}$
and we compared the model predictions with that of the SM.
One can see that,
as for the models in Figs.~\ref{fig:spe_vector_a} and~\ref{fig:spe_vector_b}
discussed above,
the effects of the light mediator are similar for the Ar and CsI detectors
and the vector boson mediator contribution increases for small values of $T_{\text{nr}} \simeq |\vec{q}|^2 / 2 M$
because of the propagator in Eq.~\eqref{Q_ll}.

In the case of the $L_\mu-L_\tau$ model the $Z'$ contribution to
$Q^{V}_{\mu,\mathrm{SM+V}}$ is positive and there can be a cancellation
with the negative SM contribution.
The cancellation occurs at
\begin{equation}
T_{\text{nr}}
=
-
\frac{1}{2M}
\left(
\frac{ \alpha_{\mathrm{EM}} g_{Z'}^2 }{ 3 \pi \sqrt{2} G_{F} }
\,
\ln\!\left(\frac{m_{\tau}^2}{m_{\mu}^2}\right)
\,
\frac{ Z F_{Z}(|\vec{q}|^{2}) }
     { g_{V}^{p} Z F_{Z}(|\vec{q}|^{2})
     + g_{V}^{n} N F_{N}(|\vec{q}|^{2}) }
+
M_{Z'}^2
\right)
,
\label{LmLt_canc}
\end{equation}
which corresponds to
$ T_{\text{nr}} \simeq 23~\text{keV} $ for Ar in Fig.~\ref{fig:spe_vector_c}
and
$ T_{\text{nr}} \simeq 6~\text{keV} $ for CsI in Fig.~\ref{fig:spe_vector_d}.
Since there is no cancellation of the SM contribution of $\nu_e$,
which does not interact with the $Z'$,
there are only shallow dips at these energies in Figs.~\ref{fig:spe_vector_c} and~\ref{fig:spe_vector_d} for this model.
The total differential rate is smaller than the SM differential rate for energies above the dip for the same reason that has been discussed above for the $B_y+L_\mu+L_\tau$ model.

In the case of the $L_e-L_\tau$ model,
there can be a cancellation of the positive $Z'$ contribution to
$Q^{V}_{e,\mathrm{SM+V}}$
with the negative SM contribution,
but it is difficult to estimate for which value of $ T_{\text{nr}} $
because of the strong dependence of $\varepsilon_{\tau e}$
on $T_{\mathrm{nr}} \simeq |\vec{q}|^2 / 2 M$
shown in Fig.~\ref{fig:loopint}.
However, one can see from Figs.~\ref{fig:spe_vector_c} and~\ref{fig:spe_vector_d}
that there are shallow dips of the differential rates at values of
$T_{\mathrm{nr}}$ that are larger than in the $L_\mu-L_\tau$ model,
because $\varepsilon_{\tau e}>\varepsilon_{\tau\mu}$,
as shown in Fig.~\ref{fig:loopint}.
The dip is more shallow than in the $L_\mu-L_\tau$ model
because the $\nu_e$ contribution to the \cenns event rate is
smaller than the sum of the $\nu_\mu$ and $\bar\nu_\mu$ contributions.

In the case of the $L_e-L_\mu$ model,
the situation is more complicated,
because the $Z'$ contribution to
$Q^{V}_{e,\mathrm{SM+V}}$
is positive,
since
$\varepsilon_{\mu e}>0$,
but the $Z'$ contribution to
$Q^{V}_{\mu,\mathrm{SM+V}}$
is negative,
since
$\varepsilon_{e\mu}<0$.
Therefore,
the $Z'$ contributions
of the dominant $\nu_\mu$ and $\bar\nu_\mu$ fluxes
enhance the \cenns differential event rate with respect to the SM prediction,
whereas the subdominant $\nu_e$ flux generate a decrease for sufficiently large values of $T_{\mathrm{nr}}$
(about 40~keV for Ar in Fig.~\ref{fig:spe_vector_c}
and
15 keV for CsI in Fig.~\ref{fig:spe_vector_d})
As a result of these opposite contributions,
the total \cenns differential rates of the $L_e-L_\mu$ model
shown in Figs.~\ref{fig:spe_vector_c} and~\ref{fig:spe_vector_d}
are only slightly larger than the SM rates in the large-$T_{\mathrm{nr}}$
parts of the figures.

\begin{figure}[!t]
\centering
\subfigure[]{
\includegraphics[width=0.48\textwidth]{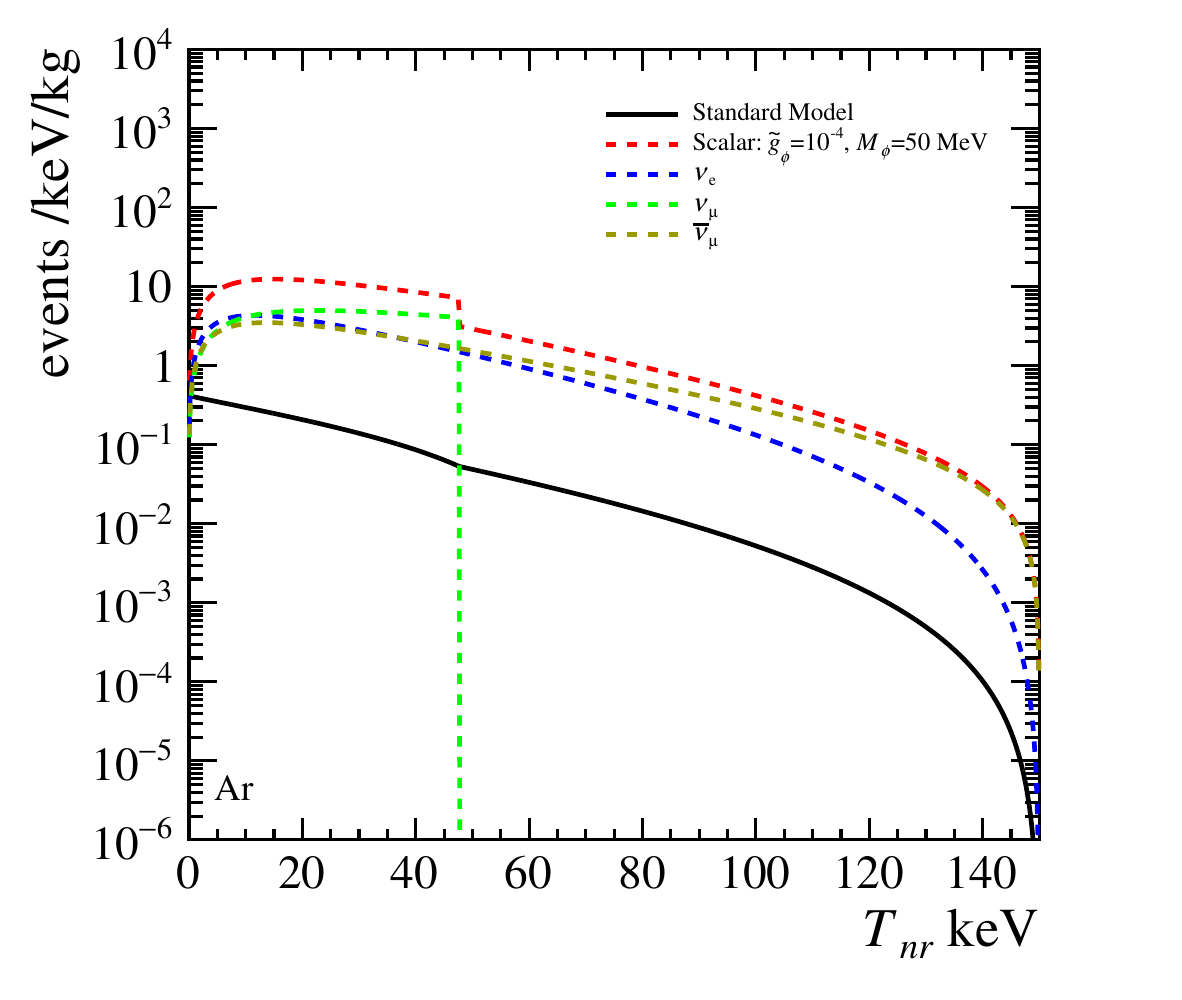}
}
\subfigure[]{
\includegraphics[width=0.48\textwidth]{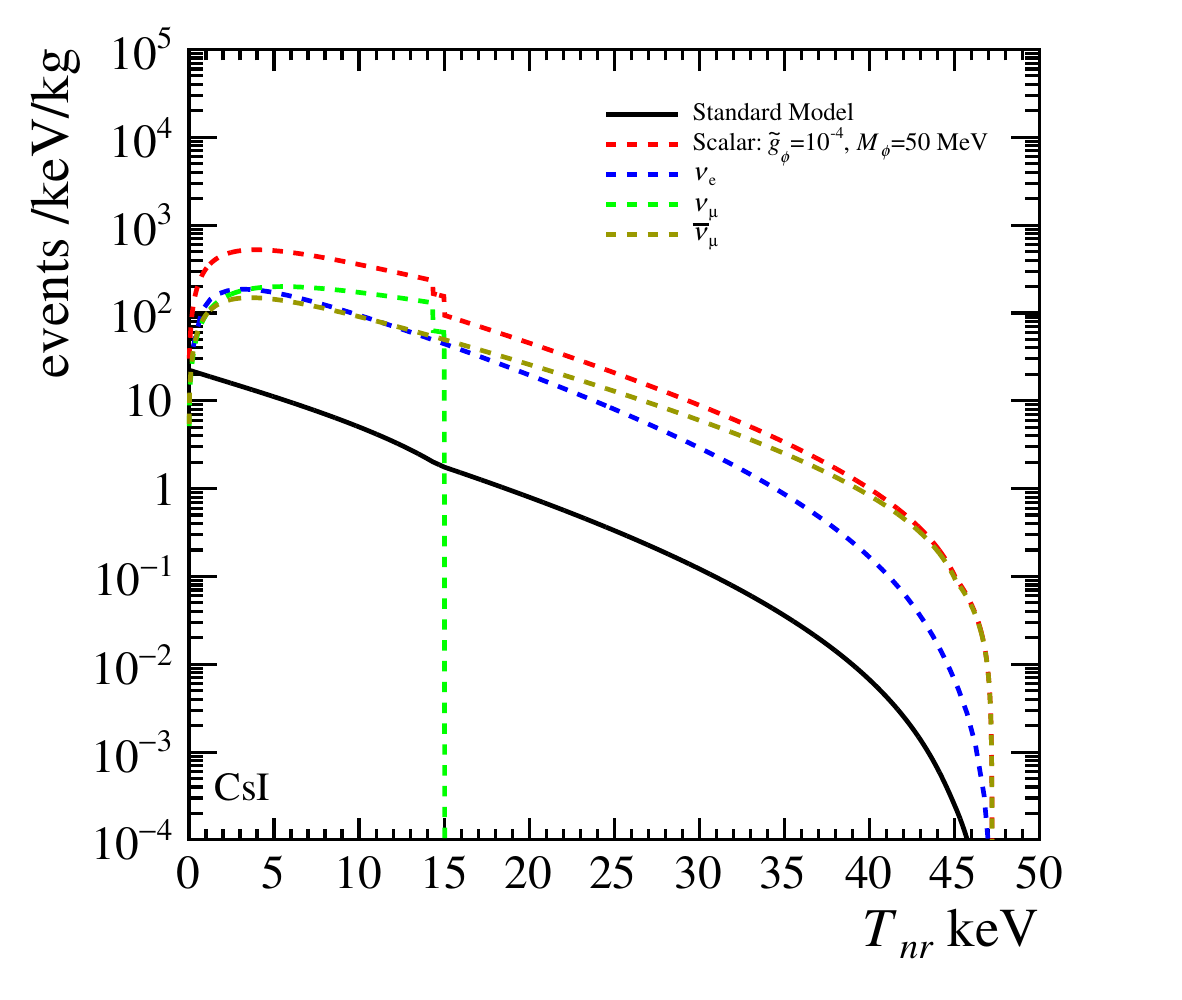}
}
\caption{Predicted CE$\nu$NS differential event rates corresponding to the experimental configuration and data taking time of the COHERENT Ar (a) and CsI (b) detectors in the universal scalar mediator model.}
\label{fig:spe_scalar}
\end{figure}

\subsection{Light scalar mediator}
\label{subs:scalar_mediator}

Non-standard neutrino interactions mediated by a scalar boson $\phi$
are possible if the SM fermion content is extended with the addition of right-handed neutrinos.
The generic Lagrangian that describes the interaction of $\phi$ with neutrinos and quarks is
\begin{equation}
\mathcal{L}_{\phi}^{S}
=
-
\phi
\left[
\sum_{\ell=e,\mu,\tau}
g_{\phi}^{\nu_{\ell}}
\,
\overline{\nu_{\ell}} \, \nu_{\ell}
+
\sum_{q=u,d}
g_{\phi}^{q}
\,
\overline{q} \, q
\right]
,
\label{Lphi}
\end{equation}
where
$\nu_{\ell} = \nu_{\ell L} + \nu_{\ell R}$
and
$g_{\phi}^{\nu_{\ell}}$ and $g_{\phi}^{q}$
are the coupling constants.
The contribution of the scalar boson interaction to the \cenns cross section
adds incoherently to the SM cross section~\cite{Lindner:2016wff,Cerdeno:2016sfi,Farzan:2018gtr,AristizabalSierra:2018eqm,AristizabalSierra:2019ykk}
\begin{equation}\label{cs_total_scalar}
	\dfrac{d\sigma_{\nu_{\ell}\text{-}\mathcal{N}}}{d T_{\text{nr}}}
	=
	\left(\dfrac{d\sigma_{\nu_{\ell}\text{-}\mathcal{N}}}{d T_{\text{nr}}}\right)_{\mathrm {SM }}
	+
	\left(\dfrac{d \sigma_{\nu_{\ell}\text{-}\mathcal{N}}}{d T_{\text{nr}}}\right)_{\mathrm {scalar }},
\end{equation} 
with
\begin{equation}\label{cs_scalar}
\left(
\frac{d \sigma_{\nu_{\ell}\text{-}\mathcal{N}}}{d T_{\text{nr}}}
\right)_{\text{scalar}}
=
\frac{ M^{2} T_{\text{nr}} }{ 4 \pi E^{2} }
\,
\frac{ (g_{\phi}^{\nu_{\ell}})^2 \mathcal{Q}_{\phi}^{2} }
     { (|\vec{q}|^{2}+M_{\phi}^{2})^{2} }
,
\end{equation}
where $\mathcal{Q}_{\phi}$ is the scalar charge of the nucleus,
given by
\begin{equation}
\mathcal{Q}_{\phi}
=
Z F_{Z}(|\vec{q}|^{2}) \sum_{q=u, d} g_{\phi}^{q} 
\langle p\vert\bar{q}q\vert p\rangle
+
N F_{N}(|\vec{q}|^{2}) \sum_{q=u, d} g_{\phi}^{q}
\langle n\vert\bar{q}q\vert n\rangle
.
\label{Q_scalar1}
\end{equation}
It is sometimes written as~\cite{Cerdeno:2016sfi,Farzan:2018gtr,AristizabalSierra:2018eqm,AristizabalSierra:2019ykk}
\begin{equation}
\mathcal{Q}_{\phi}
=
Z F_{Z}(|\vec{q}|^{2})
\sum_{q=u, d} g_{\phi}^{q} \frac{m_{p}}{m_{q}} f_{q}^{p}
+
N F_{N}(|\vec{q}|^{2})
\sum_{q=u, d} g_{\phi}^{q} \frac{m_{n}}{m_{q}} f_{q}^{n}
,
\label{Q_scalar2}
\end{equation}
with the quark contributions to the nucleon masses
\begin{equation}
f_{q}^{\mathbb{N}}
=
\frac{m_{q}}{m_{n}}
\,
\langle \mathbb{N} \vert \bar{q} q \vert \mathbb{N} \rangle
,
\label{fqN}
\end{equation}
for $\mathbb{N}=p,n$.
Since the scalar currents are not conserved,
the scalar charges of the nucleons are not simply given by the sums of the charges of their valence quarks, as in the case of a vector boson mediator
(see Eq.~\eqref{Q_ll}).
The proton and neutron matrix elements of the scalar quark current
must be calculated
(see, e.g., the recent Refs.~\cite{Hoferichter:2015dsa,Durr:2015dna,Ellis:2018dmb,Alexandrou:2019brg}).
For simplicity, we consider equal couplings
for the $u$ and $d$ quarks
and equal couplings for $\nu_e$ and $\nu_\mu$
\begin{equation}
g_{\phi}^{u}
=
g_{\phi}^{d}
=
g_{\phi}^{q}
\quad
\text{and}
\quad
g_{\phi}^{\nu_e}
=
g_{\phi}^{\nu_\mu}
=
g_{\phi}^{\nu}
.
\label{gphi}
\end{equation}
Then, we have
\begin{equation}
\mathcal{Q}_{\phi}
= 
g_{\phi}^{q}
\left[
Z F_{Z}(|\vec{q}|^{2}) 
\langle p\vert\bar{u}u+\bar{d}d\vert p\rangle
+
N F_{N}(|\vec{q}|^{2})
\langle n\vert\bar{u}u+\bar{d}d\vert n\rangle
\right]
.
\label{Q_phi2}
\end{equation}
Considering the isospin approximation, we obtain\footnote{
We neglect the small $|\vec{q}|$-dependent corrections
discussed in Ref.~\cite{Hoferichter:2020osn}.
}
\begin{equation}
\langle p\vert\bar{u}u+\bar{d}d\vert p\rangle = 
\langle n\vert\bar{u}u+\bar{d}d\vert n\rangle =
\langle N\vert\bar{u}u+\bar{d}d\vert N\rangle =
\frac{ \sigma_{\pi N} }{ \overline{m}_{ud} }
,
\label{isospin}
\end{equation}
where
$ \overline{m}_{ud} = ( m_u + m_d ) / 2$
and
$\sigma_{\pi N}$
is the pion-nucleon $\sigma$-term
that has been determined in different ways in the literature
(see the recent review in Ref.~\cite{Alarcon:2021dlz}).
Recent values have been obtained
from pionic atoms and pion-nucleon scattering~\cite{Hoferichter:2015dsa,RuizdeElvira:2017stg,Friedman:2019zhc}
and
from
lattice calculations~\cite{Durr:2015dna,Alexandrou:2019brg}.
Since there are large uncertainties on the values of
$\sigma_{\pi N}$
and
$\overline{m}_{ud}$,
we choose a reference value for
$ \sigma_{\pi N} / \overline{m}_{ud} $
given by the ratio of the central value of $\sigma_{\pi N}$
determined in Ref.~\cite{Hoferichter:2015dsa}
($\sigma_{\pi N} = 59.1 \, \text{MeV} $)
and the central PDG values~\cite{ParticleDataGroup:2020ssz}
$m_u = 2.16 \, \text{MeV} $
$m_d = 4.67 \, \text{MeV} $,
that gives
\begin{equation}
\left( \frac{ \sigma_{\pi N} }{ \overline{m}_{ud} } \right)_{\text{ref}}
=
17.3
,
\label{sigref}
\end{equation}
that allows us to write the scalar cross section \eqref{cs_scalar} as
\begin{equation}\label{cs_scalar2}
\left(
\frac{d \sigma_{\nu_{\ell}\text{-}\mathcal{N}}}{d T_{\text{nr}}}
\right)_{\text{scalar}}
=
\frac{ M^{2} T_{\text{nr}} }{ 4 \pi E^{2} }
\,
\frac{ \tilde{g}_{\phi}^4 }
     { (|\vec{q}|^{2}+M_{\phi}^{2})^{2} }
\left( \frac{ \sigma_{\pi N} }{ \overline{m}_{ud} } \right)_{\text{ref}}^2
\left[
Z F_{Z}(|\vec{q}|^{2}) 
+
N F_{N}(|\vec{q}|^{2})
\right]^2
,
\end{equation}
with
\begin{equation}
\tilde{g}_{\phi}^2
=
g_{\phi}^{\nu_{\ell}} g_{\phi}^{q}
\,
\frac{ \sigma_{\pi N} / \overline{m}_{ud} }
     { \left( \sigma_{\pi N} / \overline{m}_{ud} \right)_{\text{ref}} }
.
\label{gtilde}
\end{equation}
In this way the results of other calculations can be compared with our results
by appropriate rescaling of $\tilde{g}_{\phi}$ according with the assumptions.
We guess that $\tilde{g}_{\phi}$ is practically equal to $g_{\phi}$ in Ref.~\cite{Miranda:2020tif},
where the expression (\ref{Q_scalar2}) was used for the scalar charge of the nucleus,
with the values of the $f_{q}^{\mathbb{N}}$'s given in
Ref.~\cite{Hoferichter:2015dsa},
although the assumed values of the quark masses are not specified.
Indeed,
the values of the $f_{q}^{\mathbb{N}}$'s in Ref.~\cite{Hoferichter:2015dsa}
have been obtained from the value of $\sigma_{\pi N}$ using Eq.~(13) of Ref.~\cite{Crivellin:2013ipa},
which implies
\begin{equation}
\sum_{q=u, d} \frac{m_{p}}{m_{q}} f_{q}^{p}
=
\sum_{q=u, d} \frac{m_{n}}{m_{q}} f_{q}^{n}
=
\frac{ \sigma_{\pi N} }{ \overline{m}_{ud} }
.
\label{isospin2}
\end{equation}

On the other hand,
our approach is different from that in Refs.~\cite{Cerdeno:2016sfi,Khan:2019cvi,Suliga:2020jfa},
which considered different values for the proton and neutron matrix elements
in Eq.~\eqref{Q_phi2}:
$\langle p\vert\bar{u}u+\bar{d}d\vert p\rangle = 15.1$
and
$\langle n\vert\bar{u}u+\bar{d}d\vert n\rangle = 14$.
These values correspond to a rather large $8\%$ violation of the isospin symmetry.

Let us also note that our treatment neglected the contribution of
the strange and heavier quarks,
whose contributions to the nucleon mass have very large uncertainties
(see, e.g., Table~4 of Ref.~\cite{Cirelli:2013ufw}).
If one wants to consider them,
their contributions can be taken into account by rescaling
appropriately $\tilde{g}_{\phi}$,
assuming that the coupling of $\phi$
with all quarks is the same.

Figure~\ref{fig:spe_scalar} illustrates the effect of the scalar boson mediator
on the \cenns differential event rates
that are predicted for the COHERENT Ar and CsI detectors
for $\tilde{g}_{\phi}=10^{-4}$ and $M_{\phi}=50~\text{MeV}$.
One can see that the total \cenns rates are larger than the SM rates
for all values of $T_{\text{nr}}$, because the scalar boson cross section
adds incoherently to the SM cross section,
according to Eq.~\eqref{cs_total_scalar}.
In the two panels of Fig.\ref{fig:spe_scalar}
one can also notice that the total \cenns rates represented by the red-dashed lines
have small discontinuities at
$ T_{\text{nr}} = 47.7~\text{keV} $ for Ar
and
$ T_{\text{nr}} \simeq 15~\text{keV} $ for CsI.
These values correspond to the maximum nuclear kinetic energy
$ T_{\text{nr}}^{\text{max}} = 2 E^2 / M $
for the monoenergetic $\nu_\mu$ from pion decay ($E = 29.8~\text{MeV}$),
as shown by the green-dashed lines that represent the $\nu_\mu$ contributions.
One can see that there is an effect
also for the SM differential event rates,
which change slope at the same values of $T_{\text{nr}}$.
The effect for the scalar boson contribution is larger
because it is enhanced by the $T_{\text{nr}}$ in the numerator of
the scalar cross section, see Eq.~\eqref{cs_scalar}.
Such a dependence causes also the decrease of the scalar contribution
for very low values of $T_{\text{nr}}$ that is visible in Fig.~\ref{fig:spe_scalar}.

\section{Constraints on light mediator models}
\label{sec:result}

In this Section we present the results of the analyses of the COHERENT CsI and Ar data
with the light-mediator models described in Section~\ref{sec:cs}.
Since the data are fitted well by the SM \cenns prediction,
we obtain constraints on the mass and coupling of the light mediator in each model.
Let us note that the constraints that can be obtained with previous COHERENT CsI and Ar data
have been presented in Refs.~\cite{Liao:2017uzy,Denton:2018xmq,Dutta:2019eml,Abdullah:2018ykz,Cadeddu:2020nbr,Banerjee:2021laz}
for the more popular universal, $B-L$, and $L_\mu-L_\tau$ models
and in Ref.~\cite{Coloma:2020gfv}
for the $B-3L_e$, $B-3L_\mu$, and $B_y+L_\mu+L_\tau$ models.

In the following subsections,
we present the $2\sigma$ (95.45\% C.L.) limits obtained from the COHERENT Ar and CsI data for the models discussed in Section~\ref{sec:cs}
and we compare them with the constraints of other experiments
by using the \textbf{darkcast}~\cite{Ilten:2018crw} code for recasting the limits in the different models under consideration.
In particular,
we compare the constraints on the light vector boson mediator
obtained from the COHERENT data with the excluded regions
obtained from searches of visible dark photon decays in 
beam dump
(E141~\cite{Riordan:1987aw},
E137~\cite{Bjorken:1988as},
E774~\cite{Bross:1989mp},
KEK~\cite{Konaka:1986cb},
Orsay~\cite{Davier:1989wz,Bjorken:2009mm,Andreas:2012mt},
$\nu$-CAL~I~\cite{Blumlein:1990ay,Blumlein:1991xh,Blumlein:2011mv,Blumlein:2013cua},
CHARM~\cite{CHARM:1985anb,Gninenko:2012eq},
NOMAD~\cite{NOMAD:2001eyx},
and
PS191~\cite{Bernardi:1985ny,Gninenko:2011uv}), 
fixed target
(A1~\cite{Merkel:2014avp} and APEX~\cite{APEX:2011dww}),
collider
(BaBar~\cite{BaBar:2014zli},
KLOE~\cite{KLOE-2:2011hhj,KLOE-2:2016ydq},
LHCb~\cite{LHCb:2017trq}),
and
rare-meson-decay (NA48/2~\cite{NA482:2015wmo}) experiments,
and searches of invisible dark photons decays in the 
NA64~\cite{NA64:2019auh} 
and BaBar~\cite{BaBar:2017tiz}
experiments.
We also compare the constraints with the excluded regions obtained from  the global analysis of oscillation data (OSC)\cite{Coloma:2020gfv}.

\begin{figure}[!t]
\centering
\includegraphics[width=0.48\textwidth]{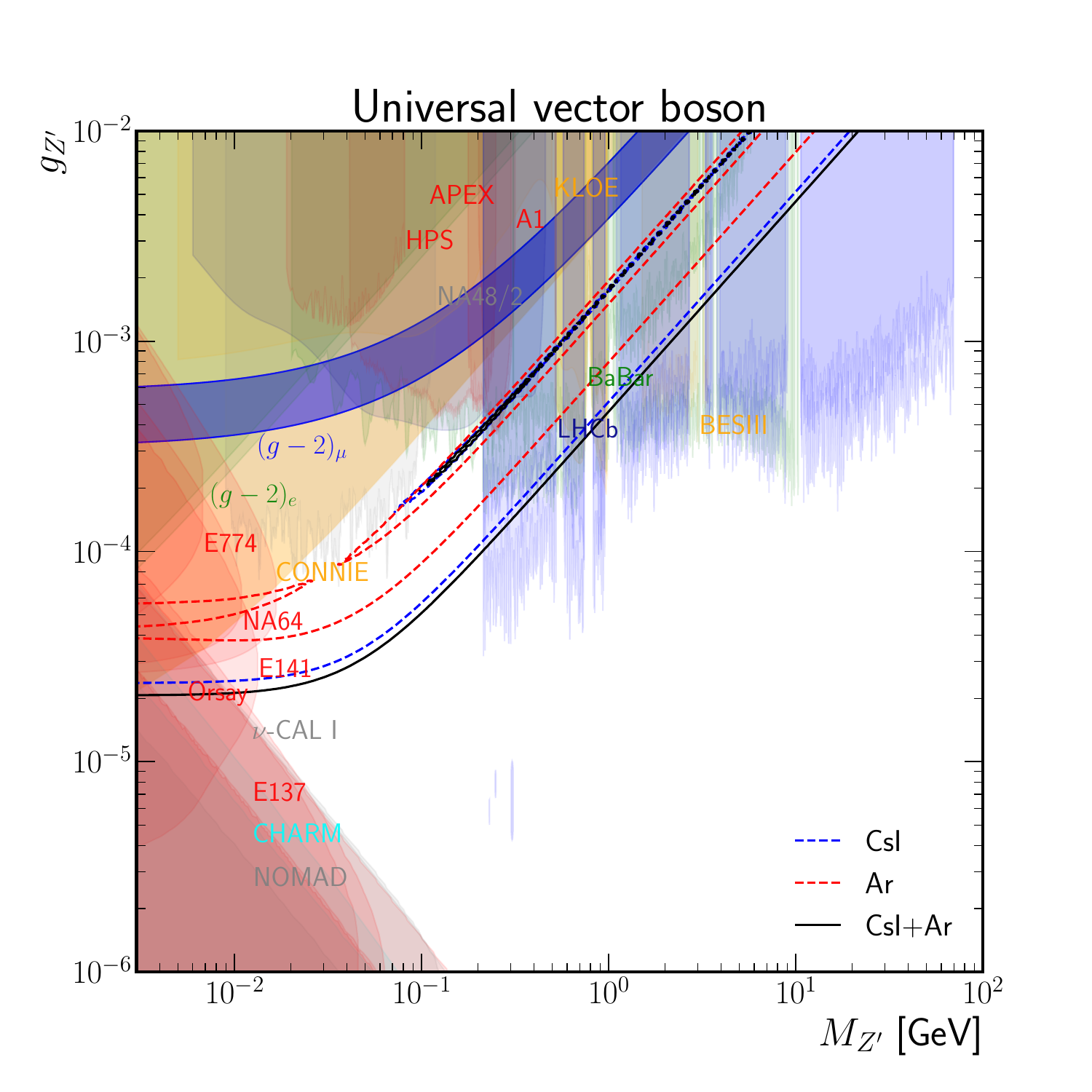}
\caption{Excluded regions (2$\sigma$) in the $M_{Z'}$-$g_{Z'}$ plane for the universal vector mediator model.}
\label{fig:universal}
\end{figure}

\subsection{Universal $Z'$ model}
\label{subs:universal}

Figure~\ref{fig:universal} shows the $2\sigma$ limits that we
obtained from the COHERENT Ar and CsI data for the universal $Z'$
model~\cite{Liao:2017uzy,Papoulias:2017qdn,Billard:2018jnl,Papoulias:2019txv,Khan:2019cvi,Cadeddu:2020nbr,Bertuzzo:2021opb}.
The black line delimits the $2\sigma$ allowed regions obtained from the combined analysis of the CsI and Ar data, while the blue and red lines delimit the excluded regions obtained from the CsI and Ar data, respectively.

Considering the combined analysis of the CsI and Ar data, one can see that in the low-mass region the black line, which represents
the upper boundary of the $2\sigma$ allowed region, flattens due to the fact that the contribution of the $Z'$ boson to $Q^{V}_{\ell,\mathrm{SM+V}}$
is small.
This happens
for $M_{Z'} \ll 100~\text{MeV}$,
because $g_{Z'}$ is small
and the boundary does not depend on $M_{Z'}$ since
$|\vec{q}| \gg M_{Z'}$ in the $Z'$ boson propagator.
On the other hand,
for higher masses the contribution of the $Z'$ boson
is suppressed by a large $M_{Z'}$,
which is dominant in the propagator,
and the boundary is given by a diagonal line proportional to $M_{Z'}$.
The numerical values of the $2\sigma$ limits in these two simple cases
are given in Table~\ref{tab:results}.

In the upper-middle part of Fig.~\ref{fig:universal},
one can see that another black line
delimits a thin diagonal strip, where
$ Q^{V}_{\ell,\mathrm{SM+V}} \simeq - Q^{V}_{\mathrm{SM}} $,
corresponding to a degeneracy with the SM cross section,
as explained in Ref.~\cite{Cadeddu:2020nbr}.
Neglecting the form factors and the small proton SM contribution,
one can find that the thin allowed strip corresponds to
\begin{equation}
(g_{Z'}^{\text{univ}})_{\text{strip}}
\simeq
\sqrt{ \frac{N}{A} \, \frac{ \sqrt{2} G_{F} M_{Z'}^2 }{ 3 } }
\simeq
1.8 \times 10^{-3} \, \frac{ M_{Z'} }{ \text{GeV} }
,
\label{univ_strip}
\end{equation}
taking into account that
$(N/A)_{\text{Ar}} \simeq (N/A)_{\text{CsI}} \simeq 0.58$.
Note that the existence of the allowed strip in the universal model
is related to the possibility to have a cancellation of the \cenns differential
event rate discussed in Section~\ref{sec:cs}
(see Eq.~\eqref{univ_canc})
because it is a consequence of the different signs of the SM and $Z'$
contributions to $ Q^{V}_{\ell,\mathrm{SM+V}} $.
Indeed,
all the models that can have a cancellation of the \cenns differential
as discussed in Section~\ref{sec:cs}
(i.e. the universal,
$B_y+L_\mu+L_\tau$,
$L_e-L_\tau$, and
$L_\mu-L_\tau$
models)
have an allowed strip, as discussed in the following.
The cancellation occurs in the excluded parameter space between the lower allowed region and the thin allowed strip for
\begin{equation}
(g_{Z'}^{\text{univ}})_{\text{canc}}
\simeq
\sqrt{ \frac{N}{A} \, \frac{ \sqrt{2} G_{F} M_{Z'}^2 }{ 6 } }
\simeq
1.3 \times 10^{-3} \, \frac{ M_{Z'} }{ \text{GeV} }
,
\label{univ_canc2}
\end{equation}
where we neglected the form factors and the small proton SM contribution.

\begin{table}[!t]
\renewcommand{\arraystretch}{1.45}
\centering
\begin{tabular}{c|cc|cc|cc}
\hline
\hline
& \multicolumn{2}{c|}{Ar} & \multicolumn{2}{c|}{CsI} & \multicolumn{2}{c}{CsI+Ar} \\
model
& $g_{Z'}$(low $M_{Z'}$) & $\dfrac{g_{Z'}}{M_{Z'}}$(high $M_{Z'}$) 
& $g_{Z'}$(low $M_{Z'}$) & $\dfrac{g_{Z'}}{M_{Z'}}$(high $M_{Z'}$) 
& $g_{Z'}$(low $M_{Z'}$) & $\dfrac{g_{Z'}}{M_{Z'}}$(high $M_{Z'}$) \\
\hline
universal	& $	3.91 	\times 10^{-5}$ & $	0.82 	\times 10^{-3}$ & $	2.36 	\times 10^{-5}$ & $	0.53 	\times 10^{-3}$ & $	2.07 	\times 10^{-5}$ & $	0.48 	\times 10^{-3}$ 	\\
$B-L$	& $	5.35 	\times 10^{-5}$ & $	1.67 	\times 10^{-3}$ & $	5.27 	\times 10^{-5}$ & $	1.00 	\times 10^{-3}$ & $	4.42 	\times 10^{-5}$ & $	0.99 	\times 10^{-3}$	\\
$B_y+L_\mu+L_\tau$	& $	10.4 	\times 10^{-5}$ & $	3.58 	\times 10^{-3}$ & $	4.97 	\times 10^{-5}$ & $	1.14 	\times 10^{-3}$ & $	4.47 	\times 10^{-5}$ & $	1.04 	\times 10^{-3}$	\\
$B-3L_e$	& $	4.91 	\times 10^{-5}$ & $	1.55 	\times 10^{-3}$ & $	5.16 	\times 10^{-5}$ & $	0.96 	\times 10^{-3}$ & $	4.34 	\times 10^{-5}$ & $	0.95 	\times 10^{-3}$	\\
$B-3L_\mu$	& $	3.45 	\times 10^{-5}$ & $	1.09 	\times 10^{-3}$ & $	3.21 	\times 10^{-5}$ & $	0.64 	\times 10^{-3}$ & $	2.76 	\times 10^{-5}$ & $	0.63 	\times 10^{-3}$	\\
$B-2L_e-L_\mu$	& $	4.62 	\times 10^{-5}$ & $	1.48 	\times 10^{-3}$ & $	4.79 	\times 10^{-5}$ & $	0.89 	\times 10^{-3}$ & $	3.95 	\times 10^{-5}$ & $	0.88 	\times 10^{-3}$	\\
$B-L_e-2L_\mu$	& $	3.97 	\times 10^{-5}$ & $	1.28 	\times 10^{-3}$ & $	3.86 	\times 10^{-5}$ & $	0.75 	\times 10^{-3}$ & $	3.26 	\times 10^{-5}$ & $	0.74 	\times 10^{-3}$	\\
\hline
$L_e-L_\mu$	& $	161 	\times 10^{-5}$ & $	54.2 	\times 10^{-3}$ & $	166 	\times 10^{-5}$ & $	36.1  	\times 10^{-3}$ & $	137 	\times 10^{-5}$ & $	34.9 	\times 10^{-3}$	\\
$L_e-L_\tau$	& $	204 	\times 10^{-5}$ & $	71.1 	\times 10^{-3}$ & $	140 	\times 10^{-5}$ & $	29.9 	\times 10^{-3}$ & $	125 	\times 10^{-5}$ & $	26.6 	\times 10^{-3}$	\\
$L_\mu-L_\tau$	& $	234 	\times 10^{-5}$ & $	80.9 	\times 10^{-3}$ & $	116 	\times 10^{-5}$ & $	26.6 	\times 10^{-3}$ & $	103 	\times 10^{-5}$ & $	24.2 	\times 10^{-3}$	\\
\hline
& $\tilde{g}_{\phi}$(low $M_\phi$) & $\dfrac{\tilde{g}_{\phi}}{M_{\phi}}$(high $M_\phi$) 
& $\tilde{g}_{\phi}$(low $M_\phi$) & $\dfrac{\tilde{g}_{\phi}}{M_{\phi}}$(high $M_\phi$) 
& $\tilde{g}_{\phi}$(low $M_\phi$) & $\dfrac{\tilde{g}_{\phi}}{M_{\phi}}$(high $M_\phi$) \\
\hline
scalar	& $	2.30 	\times 10^{-5}$ & $	0.58 	\times 10^{-3}$ & $	1.80 	\times 10^{-5}$ & $	0.31 	\times 10^{-3}$ & $	1.68 	\times 10^{-5}$ & $	0.30 	\times 10^{-3}$	\\
\hline
\hline
\end{tabular}
\caption{The $2\sigma$ (95.45\% C.L.) upper bounds on the coupling of the new boson mediator
obtained from the separate and combined analyses of the Ar and CsI COHERENT \cenns data
for low and high values of the boson mass
in the models considered in this paper.
$g_{Z'}/M_{Z'}$ and $\tilde{g}_{\phi}/M_{\phi}$ are in units of
$\mathrm{GeV}^{-1}$.
}
\label{tab:results}
\end{table}

One can see from Fig.~\ref{fig:universal} that
the limits obtained from the CsI data are stricter than those obtained from the Ar data and are close to those of the combined fit.
The limits obtained from the analysis of the Ar data
are more complicated and one can see that there are three corresponding red dashed lines in Fig.~\ref{fig:universal}.
The lowest one represents
the upper boundary of the $2\sigma$ allowed region
where the contribution of the $Z'$ boson to $Q^{V}_{\ell,\mathrm{SM+V}}$
is small,
similarly to the blue-dashed line below and the black line further below
that correspond to the CsI fit and the combined fit, respectively.
The two red-dashed lines above delimit the strip in which the Ar data are well-fitted
$ Q^{V}_{\ell,\mathrm{SM+V}} \simeq - Q^{V}_{\mathrm{SM}} $,
as discussed above for the combined fit.
However, since the Ar data are less constraining,
the strip is wider than those obtained from the CsI and combined analyses
and it extends to small values of $M_{Z'}$.

In Fig.~\ref{fig:universal} we compared the limits obtained from the
COHERENT \cenns data with those of non-\cenns experiments
and those of the CONNIE reactor \cenns experiment~\cite{CONNIE:2019xid}.
Figure~\ref{fig:universal} shows also
the $(g-2)_\mu$ $2\sigma$ allowed band which can explain the anomalous magnetic moment of the muon in this model~\cite{Muong-2:2021ojo,Jegerlehner:2009ry}
(see Appendix~\ref{app:gm2}).
One can see that the explanation of the $(g-2)_\mu$ anomaly with the universal model
is excluded by the combination of the non-\cenns exclusion limits in Fig.~\ref{fig:universal},
by the CONNIE \cenns bounds alone,
and by the COHERENT \cenns limits alone,
which confirm and extend the CONNIE \cenns bounds.
Moreover,
the COHERENT \cenns limits extend the total exclusion region by covering
a previously not-excluded area for
$ 20~\text{MeV} \lesssim M_{Z'} \lesssim 200~\text{MeV} $
and
$ 2 \times 10^{-5} \lesssim g_{Z'} \lesssim 10^{-4} $.
The new COHERENT \cenns limits are consistent with those obtained in Ref.~\cite{Cadeddu:2020nbr}
using the first COHERENT CsI data
and slightly extend the COHERENT \cenns exclusion region.

\begin{figure}[!t]
\centering
\subfigure[]{
\includegraphics[width=0.48\textwidth]{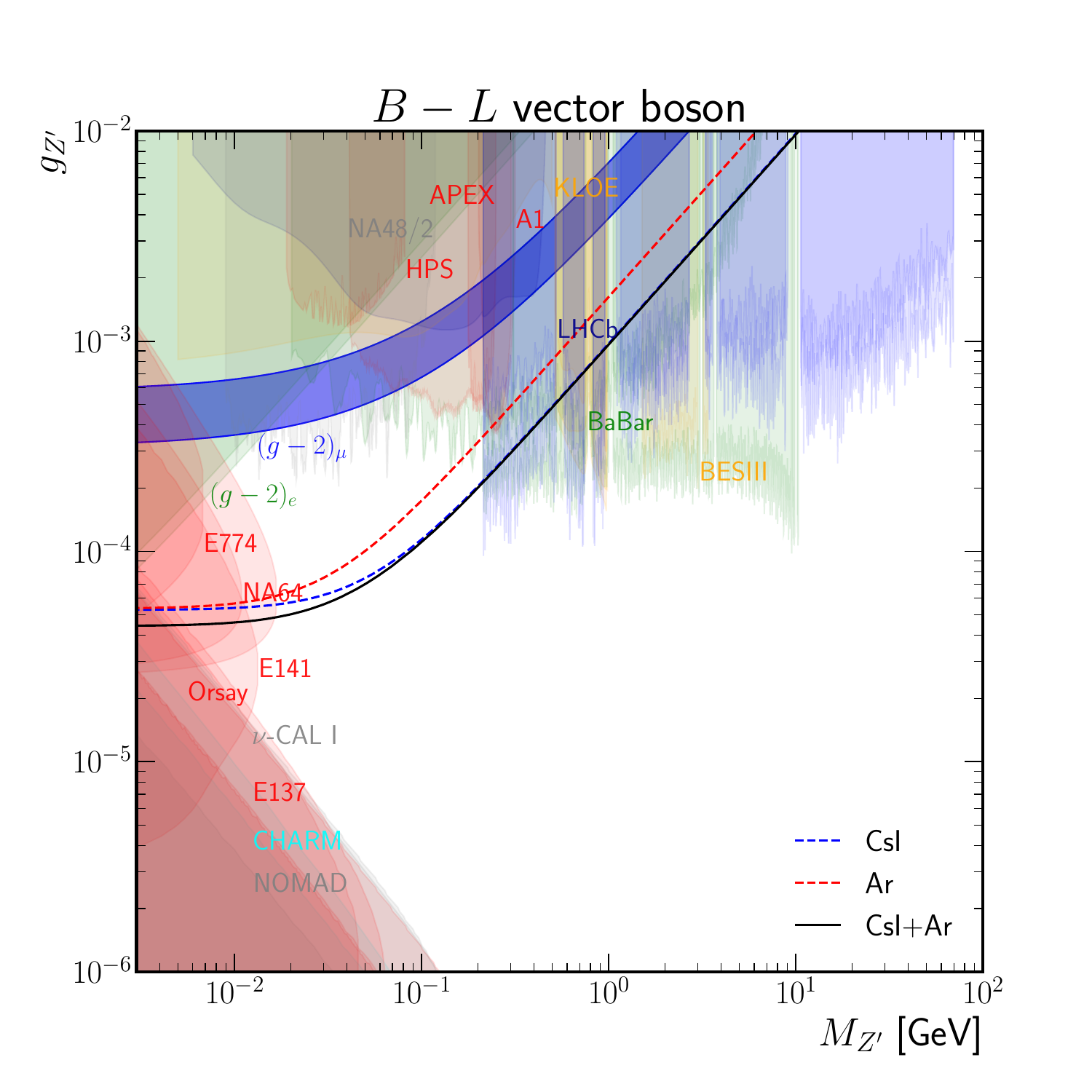}
\label{fig:B-L}
}	
\subfigure[]{
\includegraphics[width=0.48\textwidth]{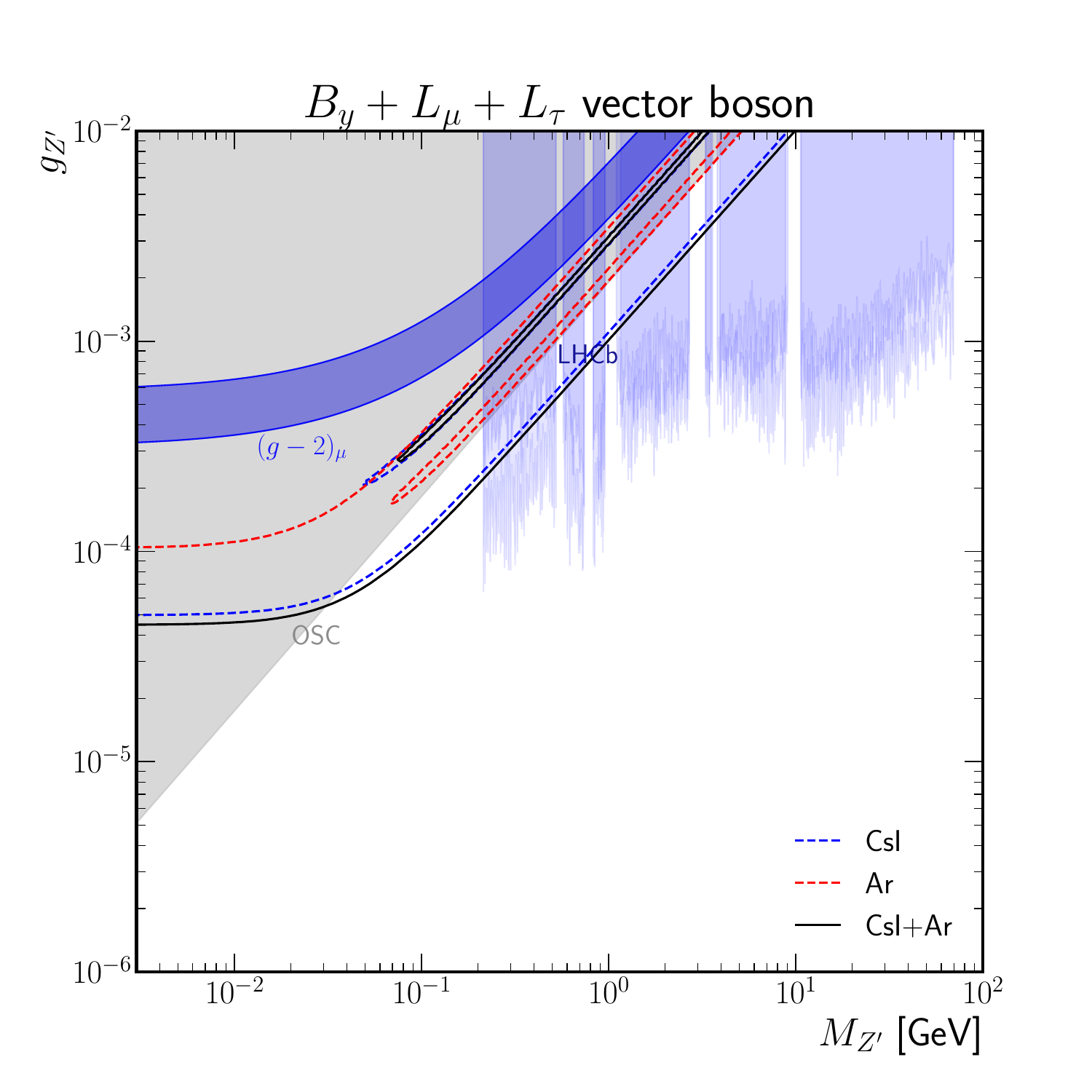}
\label{fig:By+Lmu+Ltau}
}
\caption{Excluded regions (2$\sigma$) in the $M_{Z'}$-$g_{Z'}$ plane for the
$B-L$ \subref{fig:B-L}
and
$B_y+L_\mu+L_\tau$ \subref{fig:By+Lmu+Ltau}
models.}
\label{fig:B1}
\end{figure}

\subsection{$B-L$ model}
\label{subs:B-L}

The gauged $B-L$ model is the most popular $Z'$ model
(see, e.g., the reviews in Refs.~\cite{Langacker:2008yv,Mohapatra:2014yla,Okada:2018ktp})
and its effects in \cenns have been studied in Refs.~\cite{Miranda:2020tif,Cadeddu:2020nbr,Coloma:2020gfv,delaVega:2021wpx,Bertuzzo:2021opb}
using previous COHERENT data.
Figure~\ref{fig:B-L} shows the $2\sigma$ limits that we
obtained from the COHERENT Ar and CsI data, compared with the limits obtained from other experiments
and the $(g-2)_\mu$ $2\sigma$ allowed band in this model.
One can see that the bounds obtained by experiments using only leptonic probes are the same as those for the universal model in Fig.~\ref{fig:universal},
because of the same magnitudes of the lepton charges (see Tab.~\ref{tab:charges}).
The coupling $g_{Z'}$ is well constrained by the accelerator experiments
for large values of $M_{Z'}$ and
fixed target experiments for small values of $M_{Z'}$.   
Note also that the allowed region for $(g-2)_\mu$ is the same as that in the universal model, because the magnetic moment of the muon is not dependent on the couplings of quarks.

On the other hand, the \cenns bounds are different from the universal model,
because the $Z'$ contribution to
$Q^{V}_{\mu,\mathrm{SM+V}}$
is negative and adds to the negative SM contribution.
Therefore, in Fig.~\ref{fig:B-L} there are only the upper bounds shown by the blue-dashed, red-dashed, and black-solid lines
that we obtained from the CsI, Ar, and combined analyses, respectively.
These limits have the same behaviour as the corresponding ones discussed in Subsection~\ref{subs:universal} for the universal model,
but are weaker because the quark charges are smaller by a factor of 3, as shown in Tab.~\ref{tab:charges}.
The numerical values of the limits for small and large values of $M_{Z'}$
are given in Table~\ref{tab:results}.

Figure~\ref{fig:B-L} shows that, as in the universal model,
the COHERENT \cenns limit confirms the exclusion of the explanation of the $(g-2)_\mu$ anomaly with the $B-L$ model
and
extends the total exclusion region of non-\cenns experiments by covering
a previously not-excluded area for
$ 10~\text{MeV} \lesssim M_{Z'} \lesssim 200~\text{MeV} $
and
$ 5 \times 10^{-5} \lesssim g_{Z'} \lesssim 3 \times 10^{-4} $.
Also in this case,
the new COHERENT \cenns limits are consistent with those obtained in Ref.~\cite{Cadeddu:2020nbr}
using the first COHERENT CsI data
and slightly extend the COHERENT \cenns exclusion region.

\subsection{$B_y+L_\mu+L_\tau$ model}
\label{subs:By+Lmu+Ltau}

The $2\sigma$ limits that we
obtained for $g_{Z'}$ and $M_{Z'}$ in the $B_y+L_\mu+L_\tau$
model~\cite{Farzan:2015doa,Coloma:2020gfv}
from the COHERENT Ar and CsI data are shown in Fig.~\ref{fig:By+Lmu+Ltau}.
One can see that the result of the analyses of the CsI and combined Ar and CsI data
are qualitatively similar to those discusses in Subsection~\ref{subs:universal}
for the universal model:
there is a lower curve that represents
the upper boundary of the $2\sigma$ allowed region
where the contribution of the $Z'$ boson to $Q^{V}_{\ell,\mathrm{SM+V}}$
is small
and a thin allowed strip where
$ Q^{V}_{\ell,\mathrm{SM+V}} \simeq - Q^{V}_{\mathrm{SM}} $,
leading to a degeneracy with the SM cross section that can fit well the data.
Neglecting the form factors and the small proton SM contribution,
one can find that in the case of the $B_y+L_\mu+L_\tau$ model
the thin allowed strip lies at
\begin{equation}
(g_{Z'}^{B_y+L_\mu+L_\tau})_{\text{strip}}
\simeq
\sqrt{ \frac{N}{A} \, \sqrt{2} G_{F} M_{Z'}^2 }
\simeq
3.1 \times 10^{-3} \, \frac{ M_{Z'} }{ \text{GeV} }
.
\label{By+Lmu+Ltau_strip}
\end{equation}
Under the same approximations,
one can find that the cancellation between the SM and $Z'$
contributions to $ Q^{V}_{\ell,\mathrm{SM+V}} $
occurs in the parameter space between the lower upper bound curve
and the thin allowed strip for
\begin{equation}
(g_{Z'}^{B_y+L_\mu+L_\tau})_{\text{canc}}
\simeq
\sqrt{ \frac{N}{A} \, \frac{ \sqrt{2} G_{F} M_{Z'}^2 }{ 2 } }
\simeq
2.2 \times 10^{-3} \, \frac{ M_{Z'} }{ \text{GeV} }
.
\label{By+Lmu+Ltau_canc2}
\end{equation}
Since the Ar data are less constraining than the CsI data,
the $2\sigma$ allowed region in Fig.~\ref{fig:By+Lmu+Ltau}
is that below the upper red-dashed line,
with the exception of the excluded thin strip that
corresponds to the cancellation condition, see Eq.~\eqref{By+Lmu+Ltau_canc2}.

Figure~\ref{fig:By+Lmu+Ltau} shows also the LHCb~\cite{LHCb:2017trq} limits
on $g_{Z'}$ in the $B_y+L_\mu+L_\tau$ model
and the $(g-2)_\mu$ $2\sigma$ allowed band.
One can see that the LHCb bounds exclude the $(g-2)_\mu$ allowed band
only for some ranges of values of $M_{Z'}$ above about 200~MeV.
On the other hand,
the bounds that we obtained from the analysis of the COHERENT \cenns data
exclude all the $(g-2)_\mu$ allowed band,
leading to the rejection of the
explanation of the $(g-2)_\mu$ anomaly with the $B_y+L_\mu+L_\tau$ model.

\subsection{$B-3L_e$ model}
\label{subs:B-3Le}

\begin{figure}[!t]
\centering
\subfigure[]{
\includegraphics[width=0.48\textwidth]{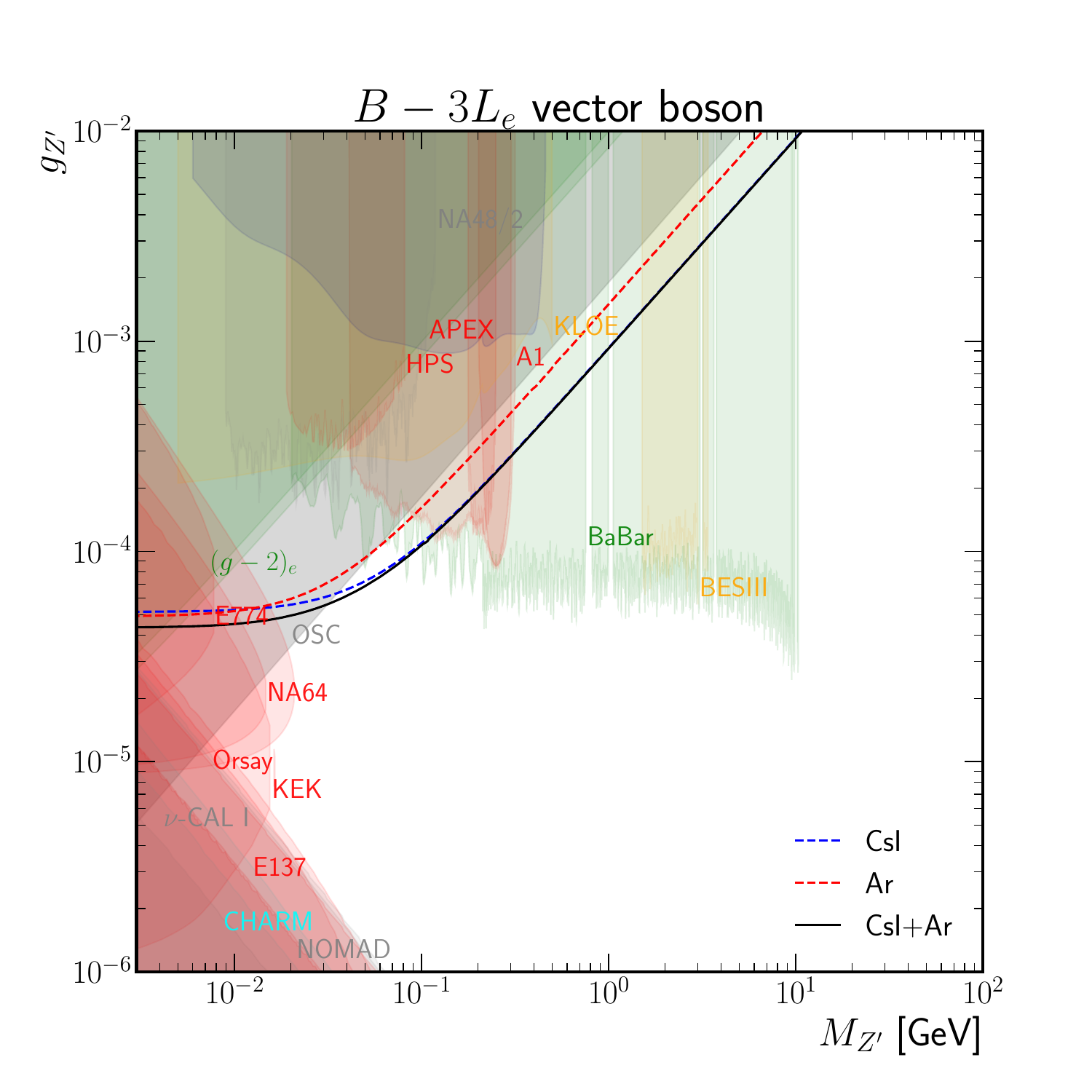}
\label{fig:B-3Le}
}
\subfigure[]{
\includegraphics[width=0.48\textwidth]{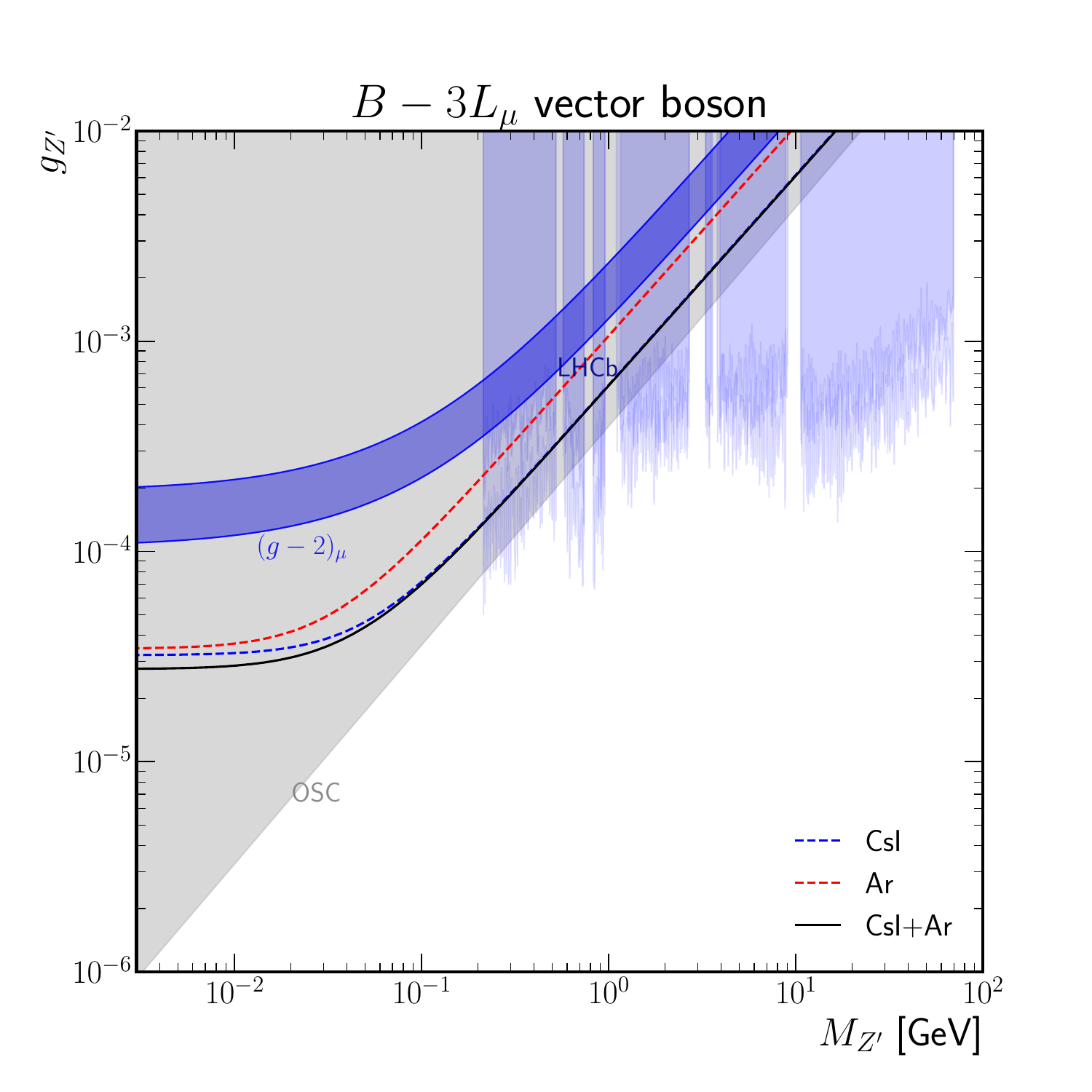}
\label{fig:B-3Lmu}
}
\caption{Excluded regions (2$\sigma$) in the $M_{Z'}$-$g_{Z'}$ plane for the
$B-3L_e$ \subref{fig:B-3Le}
and
$B-3L_\mu$ \subref{fig:B-3Lmu}
models.}
\label{fig:B2}
\end{figure}

Figure~\ref{fig:B-3Le} shows the $2\sigma$ limits that we
obtained from the COHERENT Ar and CsI data in the $B-3L_e$
model~\cite{Han:2019zkz,Heeck:2018nzc,Coloma:2020gfv,delaVega:2021wpx},
compared with the limits obtained from non-\cenns experiments,
which are quite strong,
because there are many experiments that probe the interactions of electrons
and their coupling with the $Z'$ boson in this model is three times stronger than that in the $B-L$ model.
Strict limits are especially derived from $e^{+} e^{-}$ collider data.

Note that in Fig.~\ref{fig:B-3Le} obviously there is no $(g-2)_\mu$ allowed region,
because in this model the $Z'$ boson does not interact with muonic flavor.
On the other hand, there is the $(g-2)_e$ obtained from the measurement of the magnetic moment of the electron~\cite{Hanneke:2008tm,Hanneke:2010au}
which is compatible with the prediction at $1.6 \sigma$ level taking into account the recent determination of the fine structure constant~\cite{Morel:2020dww}.

These limits that we obtained from the combined analysis of the COHERENT CsI and Ar \cenns data
have the same behaviour as the corresponding ones for the $B-L$ model.
They have also similar magnitudes,
because the lack of interaction with $Z'$ of the dominant $\nu_\mu$ and $\bar\nu_\mu$ fluxes
is compensated by the threefold increase of the $\nu_e$ coupling.
The numerical values of the limits for small and large values of $M_{Z'}$
are given in Table~\ref{tab:results}.

Figure~\ref{fig:B-3Le} shows that the COHERENT CsI and Ar \cenns data allow us to
extend the total exclusion region of non-\cenns by covering
a previously not-excluded area for
$ 10~\text{MeV} \lesssim M_{Z'} \lesssim 100~\text{MeV} $
and
$ 5 \times 10^{-5} \lesssim g_{Z'} \lesssim 2 \times 10^{-4} $.

\subsection{$B-3L_\mu$ model}
\label{subs:B-3Lmu}

Figure~\ref{fig:B-3Lmu} shows the $2\sigma$ limits that we
obtained from the COHERENT Ar and CsI data in the $B-3L_\mu$
model~\cite{Heeck:2018nzc,Coloma:2020gfv,delaVega:2021wpx},
compared with the limits obtained from the LHCb~\cite{LHCb:2017trq} experiment
($Z'\to\mu^+\mu^-$),
which exist and are relatively strong only for $M_{Z'} \gtrsim 200~\text{MeV}$.
The figure shows also the $(g-2)_\mu$ $2\sigma$ allowed band in this model,
which is not excluded by the LHCb bounds for $M_{Z'} \lesssim 200~\text{MeV}$,
but it is completely excluded by
the bounds that we obtained from the analysis of the COHERENT \cenns data.

\begin{figure}[!t]
\centering
\subfigure[]{
\includegraphics[width=0.48\textwidth]{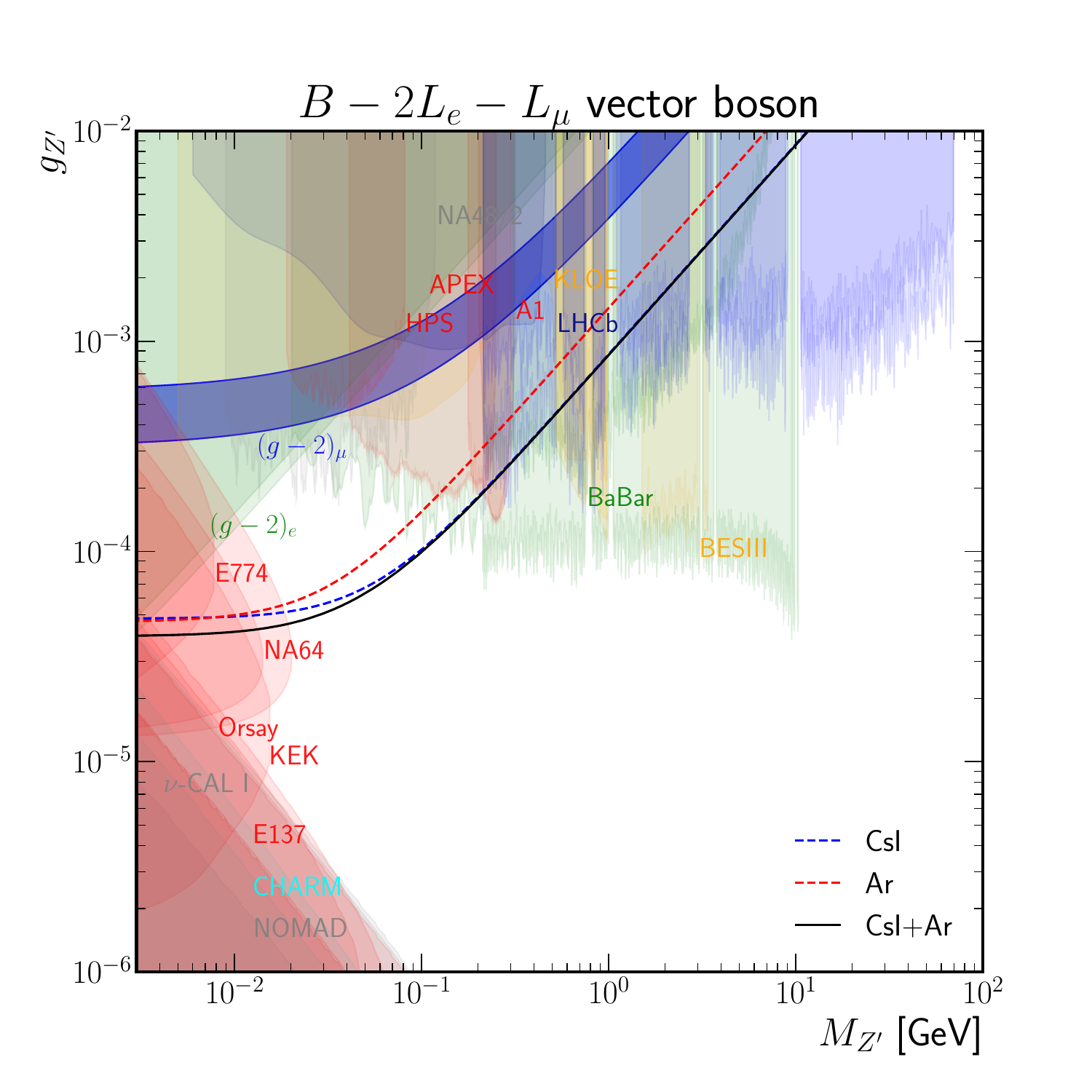}
\label{fig:B-2Le-Lmu}
}
\subfigure[]{
\includegraphics[width=0.48\textwidth]{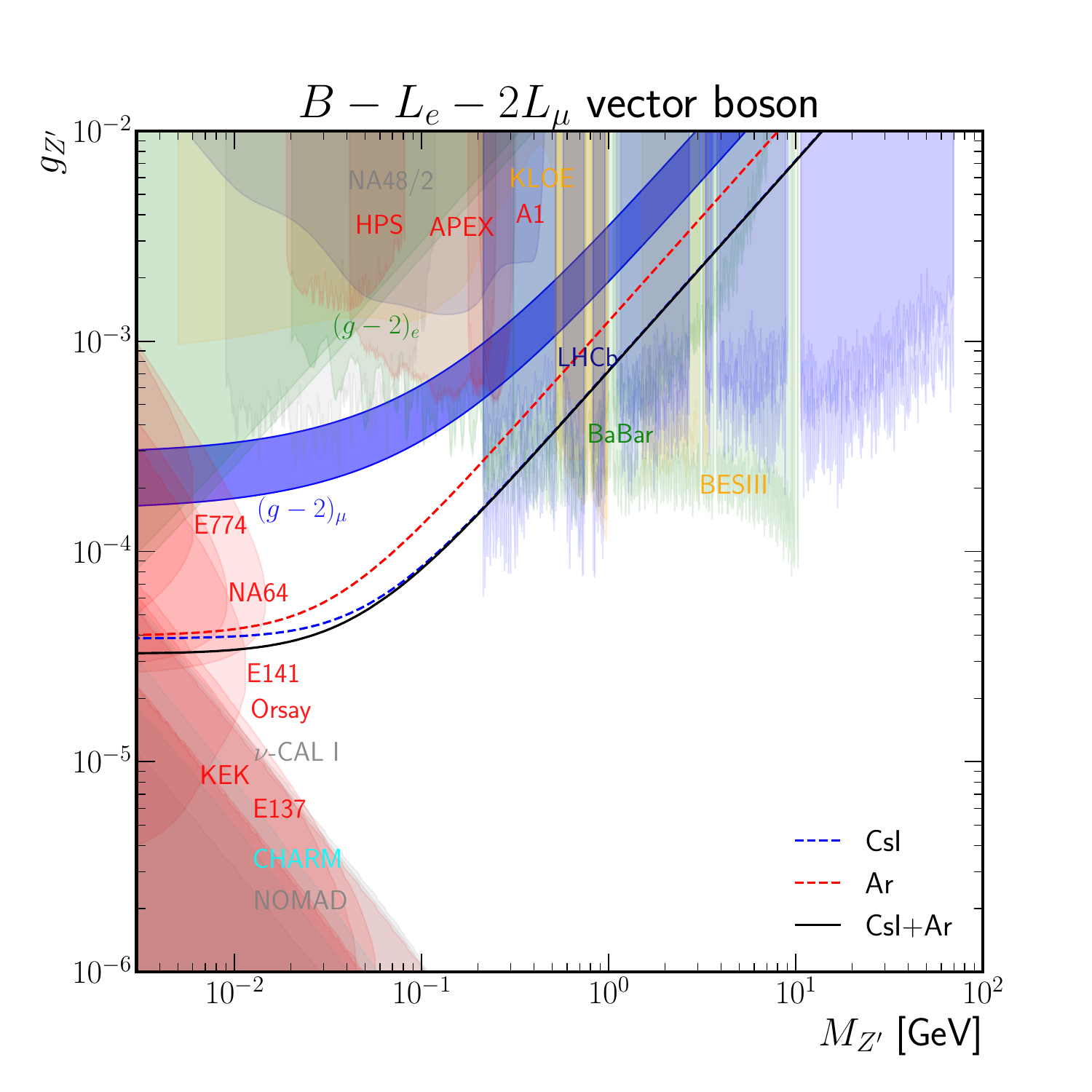}
\label{fig:B-Le-2Lmu}
}
\caption{Excluded regions (2$\sigma$) in the $M_{Z'}$-$g_{Z'}$ plane for the
$B-2L_e-L_\mu$ \subref{fig:B-2Le-Lmu}
and 
$B-L_e-2L_\mu$ \subref{fig:B-Le-2Lmu}
models.}
\label{fig:B3}
\end{figure}

\subsection{$B-2L_e-L_\mu$ model}
\label{subs:B-2Le-Lmu}

In this model~\cite{delaVega:2021wpx}
both $\nu_e$ and $\nu_\mu$ interact with the $Z'$ boson as in the $B-L$ model,
but the interaction of the subdominant $\nu_e$ flux is twice stronger.
Therefore the bounds that we obtained from the analyses of the COHERENT \cenns data,
shown in Fig.~\ref{fig:B-2Le-Lmu}
are similar and slightly stronger than those in the $B-L$ model
(see also Tab.~\ref{tab:results}).
From Fig.~\ref{fig:B-2Le-Lmu} one can also see that
the $(g-2)_\mu$ $2\sigma$ allowed band in this model
is excluded by the total exclusion limits of non-\cenns experiments.
The analysis of the COHERENT CsI and Ar \cenns data allows us to
extend the total exclusion region of non-\cenns experiments by covering
a previously not-excluded area for
$ 10~\text{MeV} \lesssim M_{Z'} \lesssim 100~\text{MeV} $
and
$ 5 \times 10^{-5} \lesssim g_{Z'} \lesssim 2 \times 10^{-4} $.

\subsection{$B-L_e-2L_\mu$ model}
\label{subs:B-Le-2Lmu}

The phenomenology of this model~\cite{delaVega:2021wpx}
is similar to that of the $B-2L_e-L_\mu$ model,
with the difference that the bounds obtained from the COHERENT \cenns data are stronger,
because the interactions with the $Z'$ boson
of the dominant $\nu_\mu$ and $\bar\nu_\mu$ fluxes are twice stronger
than those of the subdominant $\nu_e$ flux,
as one can see from Fig.~\ref{fig:B-Le-2Lmu} and Tab.~\ref{tab:results}.
One can also see from Fig.~\ref{fig:B-Le-2Lmu} that
the limits from non-\cenns are weaker
than those in Fig.~\ref{fig:B-2Le-Lmu} for the $B-2L_e-L_\mu$ model,
whereas those obtained in $\nu_\mu$ experiments are stronger.
As a result, the $(g-2)_\mu$ $2\sigma$ allowed band in this model
is not completely excluded by the results of non-\cenns experiments,
but it is completely excluded by the bounds that we obtained from
the analysis of the COHERENT CsI and Ar \cenns data.
Moreover,
we extend the total exclusion region of non-\cenns experiments by covering
a previously not-excluded area for
$ 10~\text{MeV} \lesssim M_{Z'} \lesssim 200~\text{MeV} $
and
$ 3 \times 10^{-5} \lesssim g_{Z'} \lesssim 3 \times 10^{-4} $.

\begin{figure}[!t]
\centering
\subfigure[]{
\includegraphics[width=0.48\textwidth]{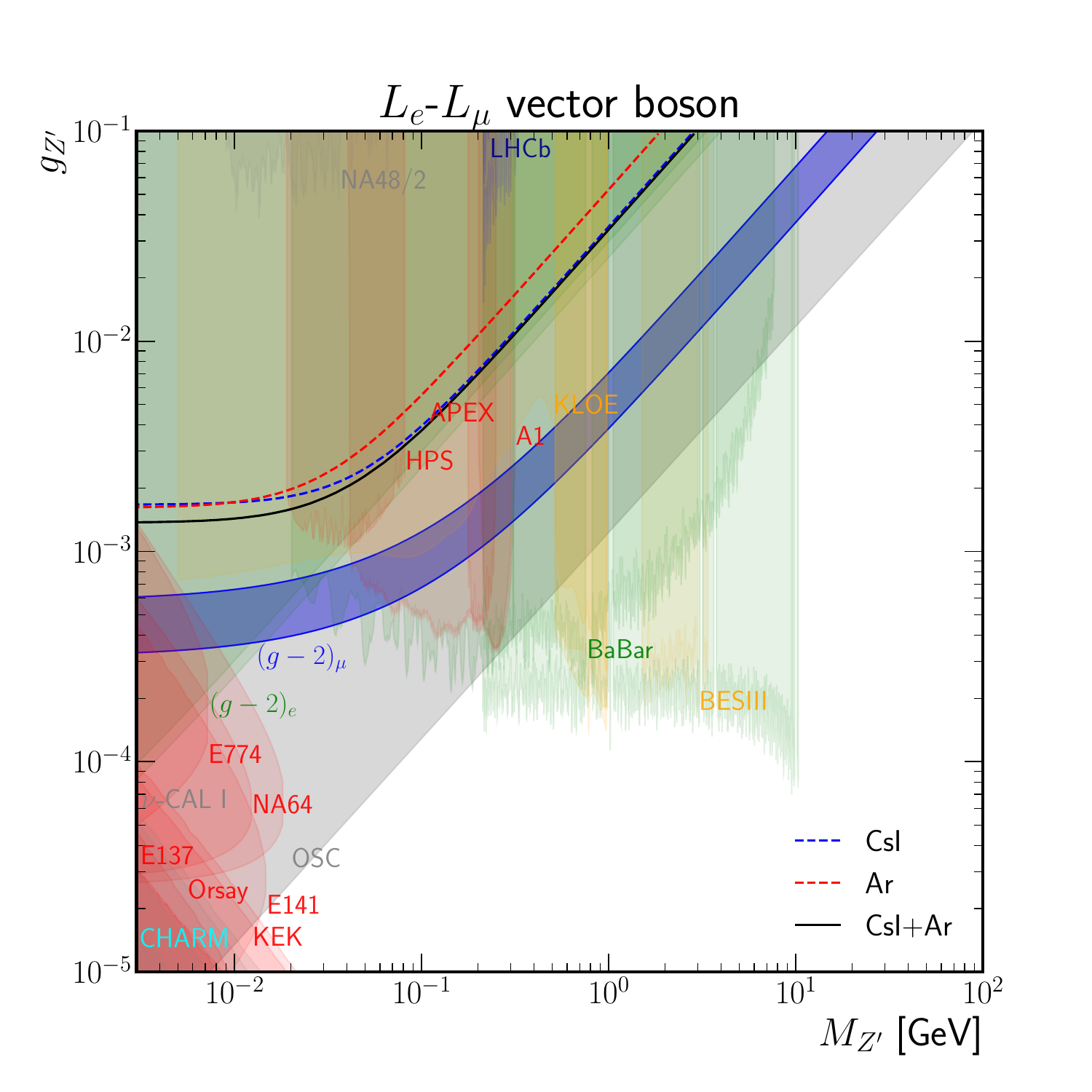}
\label{fig:Le-Lmu}
}
\subfigure[]{
\includegraphics[width=0.48\textwidth]{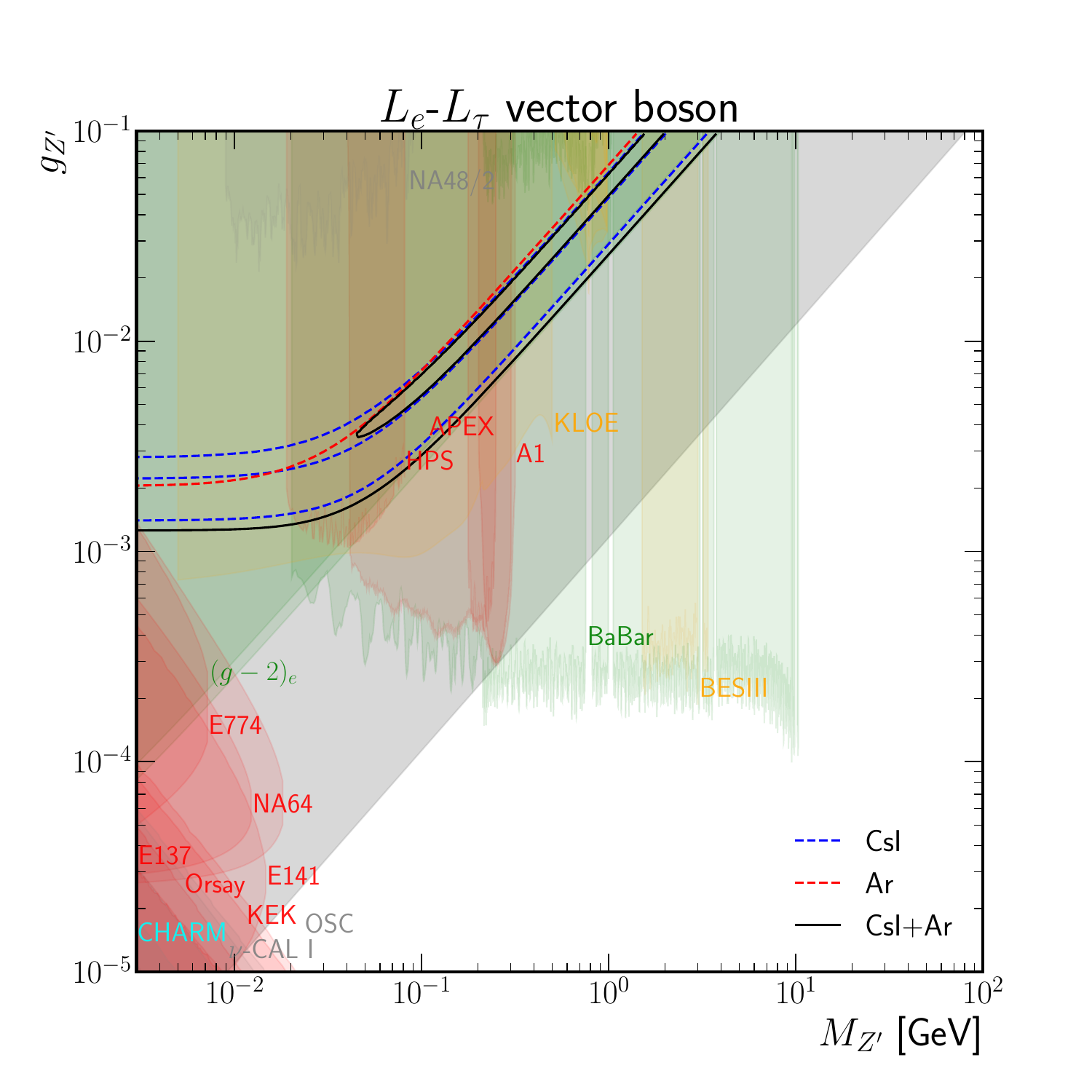}
\label{fig:Le-Ltau}
}
\subfigure[]{
\includegraphics[width=0.48\textwidth]{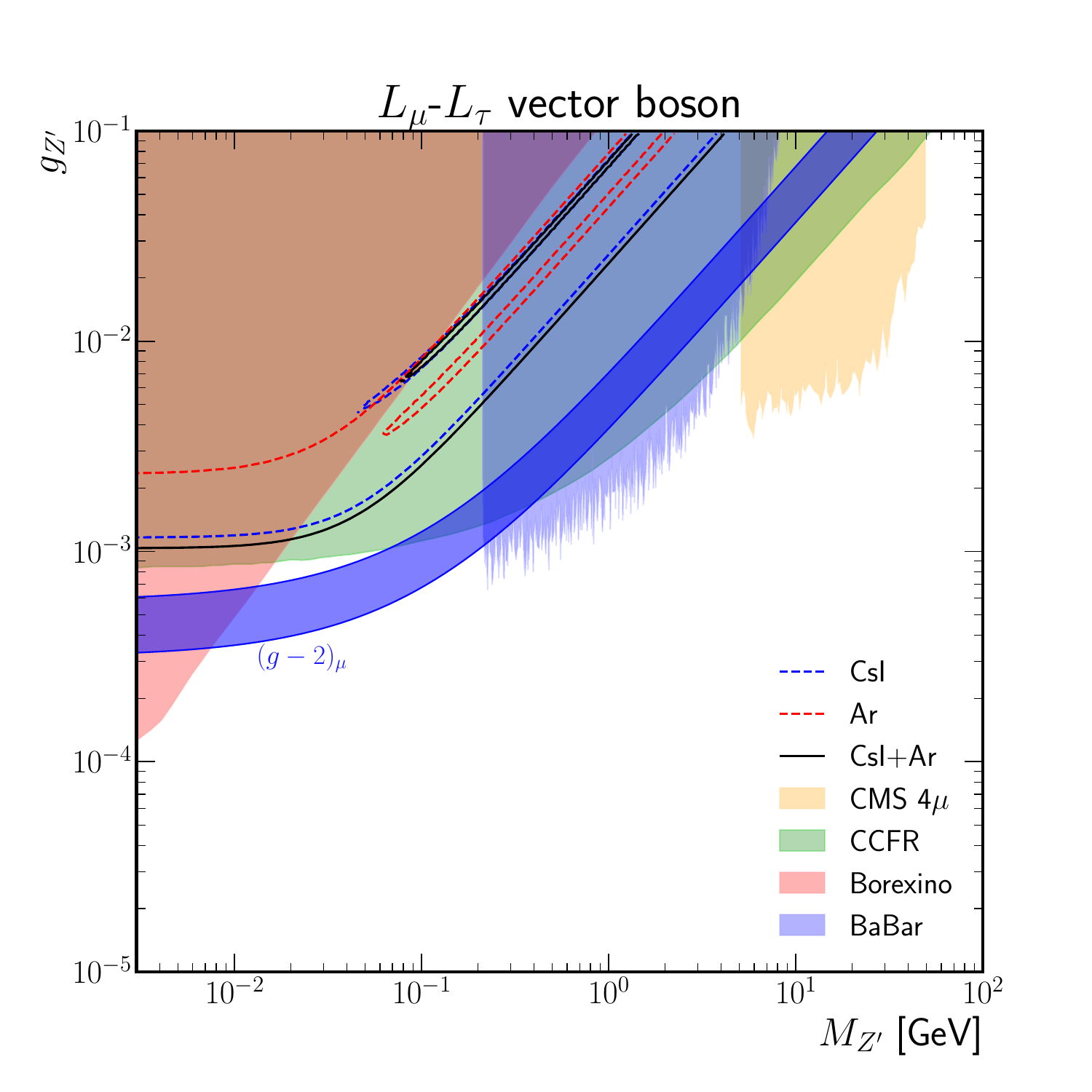}
\label{fig:Lmu-Ltau}
}
\subfigure[]{
\includegraphics[width=0.48\textwidth]{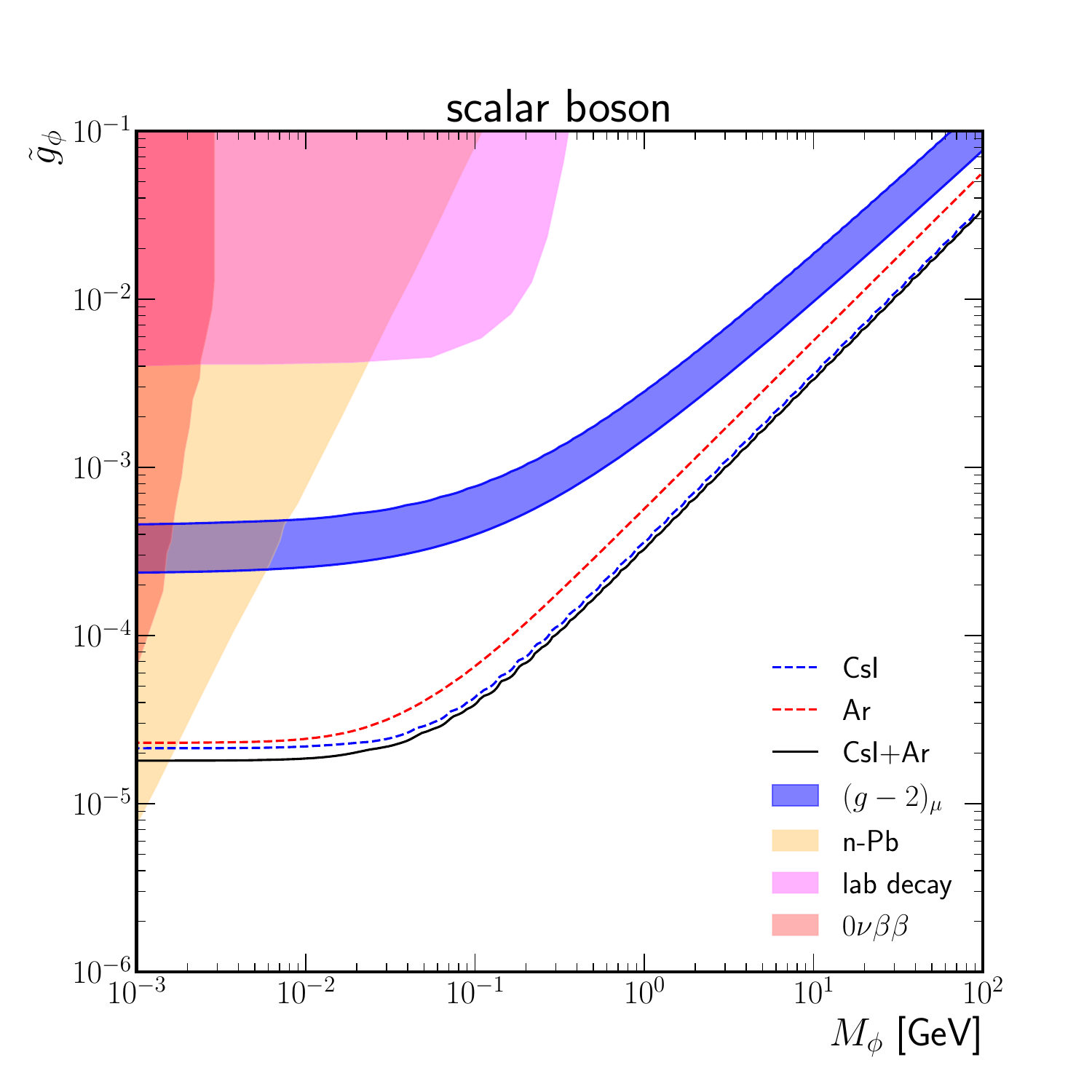}
\label{fig:scalar}
}
\caption{Excluded regions (2$\sigma$) in the mass-coupling plane for the
$L_e-L_\tau$ \subref{fig:Le-Lmu},
$L_e-L_\mu$ \subref{fig:Le-Ltau},
$L_\mu-L_\tau$ \subref{fig:Lmu-Ltau}, and
scalar \subref{fig:scalar}
models.}
\label{fig:La-Lb}
\end{figure}

\subsection{$L_e-L_\mu$ model}
\label{subs:Le-Lmu}

Figure~\ref{fig:Le-Lmu} shows the $2\sigma$ limits that we
obtained from the COHERENT Ar and CsI data in the $L_e-L_\mu$
model~\cite{He:1990pn,He:1991qd,Coloma:2020gfv}.
As for all the $L_\alpha-L_\beta$
models the constraints that we can obtain from \cenns data are weaker
than those in the previous models,
because the interaction with quarks occurs only at loop level,
and hence it is weaker.
This is also shown by the values in Table~\ref{tab:results}
where one can see that the bounds in the $L_\alpha-L_\beta$
are more than one order of magnitude weaker than those corresponding to
the models that we considered in the previous Subsections.
Moreover, in spite of the fact that all the neutrino fluxes
($\nu_e$, $\nu_\mu$, and $\bar\nu_\mu$)
interact with the $Z'$ boson in this model,
the $Z'$ contribution to the \cenns event rate is suppressed by the opposite signs
of the $\nu_e$ and $\nu_\mu$ contributions to
$Q^{V}_{\mu,\mathrm{SM+V}}$
explained at the end of Subsection~\ref{subs:vector_mediator}
and illustrated by the red-dashed curves in Figs.~\ref{fig:spe_vector_c}
and~~\ref{fig:spe_vector_d}.

One can see from Fig.~\ref{fig:Le-Lmu}
that the bounds obtained from
the current COHERENT \cenns data are not competitive with
those obtained from non-\cenns experiments
and do not contribute to the exclusion
of the $(g-2)_\mu$ $2\sigma$ allowed band in this model.
Let us note that most of this band is excluded by non-\cenns experiments,
but there is a small non-excluded part at
$M_{Z'} \approx 20-30~\text{MeV}$
and
$g_{Z'} \approx (4-7) \times 10^{-4}$.

\subsection{$L_e-L_\tau$ model}
\label{subs:Le-Ltau}

Since in the $L_e-L_\tau$ model~\cite{He:1990pn,He:1991qd,Coloma:2020gfv}
the dominant $\nu_\mu$ in the COHERENT experiment
is not interacting with the $Z'$ boson,
the bounds on the parameters of the model are rather weak.
From Fig.~\ref{fig:Le-Ltau} and Table~\ref{tab:results},
one can see that they are comparable with the bounds
in the $L_e-L_\mu$ model,
with the difference that there is an allowed diagonal strip
for $M_{Z'} \gtrsim 50~\text{MeV}$.
This occurs in the $L_e-L_\tau$ model because of the different signs of the SM
and $Z'$ contributions to $Q^{V}_{\ell,\mathrm{SM+V}}$
discussed in Subsection~\ref{subs:vector_mediator}.
The allowed strip is the region of the parameters where
$ Q^{V}_{\ell,\mathrm{SM+V}} \simeq - Q^{V}_{\mathrm{SM}} $,
leading to a degeneracy with the SM cross section,
as for the similar strips in
Fig.~\ref{fig:universal} for the universal model
and
Fig.~\ref{fig:By+Lmu+Ltau} for the $B_y+L_\mu+L_\tau$ model.
Neglecting the form factors and the small proton SM contribution,
this degeneracy occurs for
\begin{equation}
(g_{Z'}^{L_e-L_\tau})_{\text{strip}}
\approx
\sqrt{ \frac{N}{Z} \,
\frac{ \pi G_{F} M_{Z'}^2 }
     { \sqrt{2} \alpha_{\mathrm{EM}} \varepsilon_{\tau e} } }
\approx
6 \times 10^{-2} \, \frac{ M_{Z'} }{ \text{GeV} }
,
\label{Le-Ltau_strip}
\end{equation}
where we considered $N/Z \approx 1.3$
and $\varepsilon_{\tau e} \approx 1.5$.
One can see from Fig.~\ref{fig:Le-Ltau} that the allowed diagonal strip
lies along the line given by Eq.~\eqref{Le-Ltau_strip}.

The non-\cenns bounds in Fig.~\ref{fig:Le-Ltau}
are the same as the bounds in Fig.~\ref{fig:Le-Lmu}
that have been obtained with electron-interaction experiments,
including that from $(g-2)_e$
value~\cite{Hanneke:2008tm,Hanneke:2010au,Morel:2020dww}
that we already mentioned above in Subsection~\ref{subs:B-3Le}
for the $B-3L_e$ model.
From Fig.~\ref{fig:Le-Ltau}
one can see that in the $L_e-L_\tau$ model the bounds obtained from the
current COHERENT \cenns data are not competitive with
those obtained from non-\cenns experiments
and the non-\cenns experiments exclude the \cenns allowed diagonal strip
discussed above.

\subsection{$L_\mu-L_\tau$ model}
\label{subs:Lmu-Ltau}

Figure~\ref{fig:Lmu-Ltau} shows the $2\sigma$ limits that we
obtained from the COHERENT Ar and CsI data in the popular $L_\mu-L_\tau$
model~\cite{He:1990pn,Baek:2001kca,Altmannshofer:2019zhy,Banerjee:2018mnw,Gninenko:2020xys,Cadeddu:2020nbr,Banerjee:2021laz,Bertuzzo:2021opb}.
From Fig.~\ref{fig:Lmu-Ltau} and the values in Table~\ref{tab:results},
one can see that the bounds obtained in this model from the COHERENT \cenns data
are the strongest among the $L_\alpha-L_\beta$ models.
This is due to the interaction with the $Z'$ boson of the dominant
$\nu_\mu$ and $\bar\nu_\mu$
fluxes that is not suppressed by the opposite contribution of the $\nu_e$
flux as in the $L_e-L_\mu$ model.

From Fig.~\ref{fig:Lmu-Ltau},
one can also see that there is an allowed diagonal strip
that is the region of the parameters where
$ Q^{V}_{\ell,\mathrm{SM+V}} \simeq - Q^{V}_{\mathrm{SM}} $,
which is due to the different signs of the SM and $Z'$
contributions to $Q^{V}_{\ell,\mathrm{SM+V}}$,
as discussed above for other models.
Since $\varepsilon_{\tau\mu}\simeq\ln(m_\tau^2/m_\mu^2)/6$,
as discussed in Subsection~\ref{subs:vector_mediator},
the allowed diagonal strip corresponds to
\begin{equation}
(g_{Z'}^{L_\mu-L_\tau})_{\text{strip}}
\approx
\sqrt{ \frac{N}{Z} \,
\frac{ 6 \pi G_{F} M_{Z'}^2 }
     { \sqrt{2} \alpha_{\mathrm{EM}} \ln(m_\tau^2/m_\mu^2) } }
\approx
7 \times 10^{-2} \, \frac{ M_{Z'} }{ \text{GeV} }
,
\label{Lmu-Ltau_strip}
\end{equation}
where we considered $N/Z \approx 1.3$

One can see form Fig.~\ref{fig:Lmu-Ltau}
that in the $L_\mu-L_\tau$ model there are several non-\cenns constraints
whose combination is more stringent
than those given by the current COHERENT \cenns data:
CMS~\cite{CMS:2018yxg} ($Z \to Z' \mu\mu \to 4\mu$),
BaBar~\cite{BaBar:2016sci} ($e^+ e^- \to Z' \mu\mu \to 4\mu$),
CCFR~\cite{CCFR:1991lpl,Altmannshofer:2014pba} (neutrino trident production),
and
Borexino~\cite{Bellini:2011rx,Kamada:2015era,Gninenko:2020xys}
($Z'$-mediated solar neutrino interactions).
These non-\cenns constraints exclude
the allowed diagonal strip corresponding to Eq.~\eqref{Lmu-Ltau_strip}.
On the other hand, they do not completely exclude
the $(g-2)_\mu$ $2\sigma$ allowed band in this model,
that is shown in Fig.~\ref{fig:Lmu-Ltau}.
One can see that the part of this band for
$ 10~\text{MeV} \lesssim M_{Z'} \lesssim 200~\text{MeV} $
and
$ 3 \times 10^{-4} \lesssim g_{Z'} \lesssim 10^{-3} $
eludes the exclusions.

\subsection{Scalar model}
\label{subs:scalar_model}

Figure~\ref{fig:scalar} shows the $2\sigma$ limits that we
obtained from the COHERENT Ar and CsI data
in the scalar boson mediator model described
in Subsection~\ref{subs:scalar_mediator}.
The figure shows also the $(g-2)_\mu$ $2\sigma$ allowed band in this model
and the constraints obtained 
from the measurement of neutrons scattering on a ${}^{208}$Pb target~\cite{Barbieri:1975xy,Schmiedmayer:1988bm,Leeb:1992qf}, 
the measurement of $\tau$, mesons, and $Z$ decays~\cite{Berryman:2018ogk,Bilenky:1999dn,Lessa:2007up,Pasquini:2015fjv,Krnjaic:2019rsv,Brdar:2020nbj}, 
and double-beta decay experiments~\cite{Agostini:2015nwa,KamLAND-Zen:2012uen,Berryman:2018ogk,Blum:2018ljv}
(see also the summary in Ref.\cite{Suliga:2020jfa}).

One can see from Fig.~\ref{fig:scalar} that the COHERENT \cenns constraints are much more stringent
than the non-\cenns bounds for $M_\phi \gtrsim 2~\text{MeV}$
and they exclude the explanation of the $(g-2)_\mu$ anomaly
in the scalar boson mediator model.

\section{Conclusions}
\label{sec:conclusions}

In this paper we analyzed the recent \cenns data obtained by the COHERENT Collaboration with the CsI and Ar detectors and we derived constraints on the coupling and mass of a non-standard light vector or scalar boson mediator considering several models that have been studied in the literature.
We presented the results obtained from the separate analyses of the CsI and Ar data and those obtained from the combined analysis of the two datasets.

We considered several models with a light vector boson $Z'$:
the anomalous model with universal coupling of the $Z'$ vector boson with all SM fermions (assuming that the quantum anomalies are canceled by the contributions of the non-standard fermions of an extended full
theory),
several anomaly-free models with gauged $U(1)'$ symmetries, as the popular $B-L$ symmetry, in which the $Z'$ vector boson couples directly to quarks and leptons,
and the anomaly-free models with gauged
$L_e-L_\mu$,
$L_e-L_\tau$, and
$L_\mu-L_\tau$
symmetries,
in which the $Z'$ vector boson couples directly to the involved lepton flavors and indirectly to nucleons at the one-loop level.

We compared the constraints obtained from the COHERENT CsI and Ar \cenns data with those obtained from several non-\cenns experiments.
We showed that the COHERENT \cenns data allow us to extend the
excluded regions of the parameters in
the models in which the $Z'$ vector boson couples directly to quarks and
in the universal scalar mediator model.
In particular,
the total excluded region is extended to smaller values of the coupling constant
$g_{Z'}$ for
$ 10~\text{MeV} \lesssim M_{Z'} \lesssim 100~\text{MeV} $
in the universal,
$B-L$,
$B-3L_e$,
$B-2L_e-L_\mu$, and
$B-L_e-2L_\mu$
models.
The regions in the $M_{Z'}$-$g_{Z'}$ plane
that are excluded by non-\cenns experiments for the
$B_y+L_\mu+L_\tau$
and
$B-3L_\mu$
models are limited to $M_{Z'} \gtrsim 200~\text{MeV}$.
Therefore, for these models the COHERENT \cenns data
allow us to obtain a large extension of the total excluded region
for $M_{Z'} \lesssim 200~\text{MeV}$.

The models in which the $Z'$ couples to muons
can explain the $(g-2)_\mu$
anomaly~\cite{Muong-2:2006rrc,Aoyama:2020ynm,Muong-2:2021ojo},
and the allowed band in the $M_{Z'}$-$g_{Z'}$ plane
is tested by non-\cenns experiments,
as shown in Figs.~\ref{fig:universal}, \ref{fig:B1}, \ref{fig:B2}, and~\ref{fig:B3}.
The results of our analysis of the COHERENT \cenns data
exclude the explanation of the $(g-2)_\mu$
anomaly in the models in which the $Z'$ vector boson couples directly to quarks
by confirming the excluded regions of non-\cenns experiments
and extending the coverage of the $(g-2)_\mu$ allowed band
for the
$B_y+L_\mu+L_\tau$,
$B-3L_\mu$, and
$B-L_e-2L_\mu$
models.

The constraints that we obtained for the
$L_e-L_\mu$,
$L_e-L_\tau$, and
$L_\mu-L_\tau$
are less stringent because the one-loop interactions of the $Z'$ vector boson
with the nucleons is weaker than the direct interaction.
For these models the current COHERENT \cenns data
allow us to confirm the exclusion of part of the parameter space that
is already covered by non-\cenns experiments,
but cannot probe the $(g-2)_\mu$ allowed band
in the
$L_e-L_\mu$
and
$L_\mu-L_\tau$
models.

We finally considered \cenns interactions mediated by a light scalar boson $\phi$ assuming for simplicity a universal coupling
with the quarks and neutrinos involved in the \cenns processes measured in the COHERENT experiment.
We obtained the strong constraints on the mass $M_\phi$ and coupling
of the scalar boson shown in Fig.~\ref{fig:scalar}
that greatly extend the region excluded by non-\cenns experiments
and rejects the explanation of the $(g-2)_\mu$ anomaly in this model.

\acknowledgments

We would like to thank A. Konovalov and D. Pershey for the useful information provided for the analysis of the COHERENT data.
The work of C. Giunti  and C.A. Ternes is supported by the research grant "The Dark Universe: A Synergic Multimessenger Approach" number 2017X7X85K under the program PRIN 2017 funded by the Ministero dell'Istruzione, Universit\`a e della Ricerca (MIUR).
The work of Y.F. Li and Y.Y. Zhang is supported by the National Natural Science Foundation of China under Grant No.~12075255, ~12075254 and No.~11835013, by the Key Research Program of the Chinese Academy of Sciences under Grant No.~XDPB15.
The work of Y.Y. Zhang is also supported by China Postdoctoral Science Foundation under Grant No.~2021T140669.


\appendix

\begin{figure}[t!]
\centering
\subfigure[]{\label{fig:zsvec}%
\includegraphics[width=0.3\linewidth]{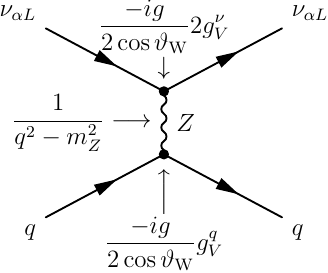}
}
\hspace{2cm}
\subfigure[]{\label{fig:zpvec}%
\raisebox{5mm}{\includegraphics[width=0.3\linewidth]{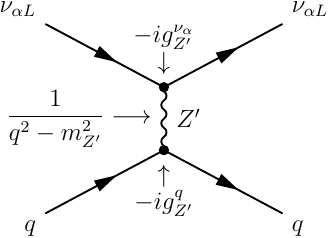}}
}
\caption{\label{fig:diagrams}
Feynman diagrams describing
\subref{fig:zsvec}
the vector part of the Standard Model neutral-current interaction
and
\subref{fig:zpvec}
the $Z'$ vector interaction
of left-handed neutrinos with quarks.
}
\end{figure}

\section{$Z'$ coupling}
\label{app:coupling}

There is some confusion on the value of the coefficient of the contribution of a new $Z'$ vector boson mediator in Eq.~\eqref{Q_ll} that is obtained assuming the interaction Lagrangian in Eq.~\eqref{LZp}.
For example, in Refs.~\cite{Liao:2017uzy,Papoulias:2017qdn,Papoulias:2018uzy,Papoulias:2019xaw,Khan:2019cvi,Dutta:2019eml}
the coefficient is half of that in Eq.~\eqref{Q_ll}.
On the other hand,
the coefficient in Refs.~\cite{Bertuzzo:2017tuf,Abdullah:2018ykz,Billard:2018jnl,Han:2019zkz,Flores:2020lji,Papoulias:2019txv,Miranda:2020tif,Cadeddu:2020nbr,delaVega:2021wpx,Bertuzzo:2021opb,CONUS:2021dwh} agrees with that in Eq.~\eqref{Q_ll}.
In this Appendix we prove that the coefficient in Eq.~\eqref{Q_ll} is the right one.

Let us start by considering the relevant vector part of the Standard Model neutral-current weak interaction Lagrangian
(see, e.g., Refs.~\cite{Giunti:2007ry,ParticleDataGroup:2020ssz})
\begin{equation}
\mathcal{L}_{Z}^{V}
=
-
\frac{ g }{ 2 \cos\vartheta_{\text{W}} }
\,
Z_{\mu}
\left[
2 g_{V}^{\nu}
\sum_{\ell=e,\mu,\tau}
\overline{\nu_{\ell L}} \gamma^{\mu} \nu_{\ell L}
+
\sum_{q=u,d}
g_{V}^{q}
\,
\overline{q} \gamma^{\mu}  q
\right]
,
\label{LNC}
\end{equation}
with the tree-level couplings
\begin{equation}
g_{V}^{\nu} = \frac{1}{2}
,
\quad
g_{V}^{u} = \frac{1}{2} - \frac{4}{3} \, \sin^2\vartheta_{\text{W}}
,
\quad
\text{and}
\quad
g_{V}^{d} = - \frac{1}{2} + \frac{2}{3} \, \sin^2\vartheta_{\text{W}}
.
\label{gNC}
\end{equation}
Confronting Eq.~\eqref{LNC} with the Lagrangian~\eqref{LZp},
one can see that
the $Z'$ vector interaction of left-handed neutrinos with quarks
is o++++++++btained from the vector part of the Standard Model neutral-current interaction
with the substitutions
\begin{equation}
\frac{ g }{ 2 \cos\vartheta_{\text{W}} }
\,
2 g_{V}^{\nu}
\to
g_{Z'}^{\nu_{\ell}V}
,
\quad
\frac{ g }{ 2 \cos\vartheta_{\text{W}} }
\,
g_{V}^{q}
\to
g_{Z'}^{q V}
,
\quad
\text{and}
\quad
m_{Z} \to m_{Z'}
.
\label{subs}
\end{equation}
This correspondence is shown in Fig.~\ref{fig:diagrams}, where we depicted the
two Feynman diagrams that describe the neutrino-quarks interactions that contribute to \cenns at tree level.
The total amplitude is given by the sum of the two diagrams
\begin{equation}
A
\propto
\frac{ g^2 }{ 4 \cos^2\vartheta_{\text{W}} }
\,
\frac{2 g_{V}^{\nu} g_{V}^{q}}{q^2-m_{Z}^2}
+
\frac{g_{Z'}^{\nu_{\ell}V} g_{Z'}^{q V}}{q^2-m_{Z'}^2}
.
\label{sub1}
\end{equation}
Taking into account that $g_{V}^{\nu}=1/2$ and
\begin{equation}
\frac{g^{2}}{4\cos^{2}\vartheta_{\text{W}} m_{Z}^{2}}
=
\sqrt{2} G_{\text{F}}
,
\label{c254}
\end{equation}
for $ q^2 \ll m_{Z}^2 $
we obtain
\begin{equation}
A
\propto
g_{V}^{q}
+
\frac{g_{Z'}^{\nu_{\ell}V} g_{Z'}^{q V}}{\sqrt{2} G_{\text{F}}\left(q^2-m_{Z'}^2\right)}
.
\label{sub2}
\end{equation}
This relation leads to Eq.~\eqref{Q_ll},
taking into account that the conservation of the vector current
implies that
\begin{equation}
g_{Z'}^{p} = 2 g_{Z'}^{u V} + g_{Z'}^{d V}
\quad
\text{and}
\quad
g_{Z'}^{n} = g_{Z'}^{u V} + 2 g_{Z'}^{d V}
.
\label{CVC}
\end{equation}

In conclusion of this Appendix,
let us note that the results of the analyses in Refs.~\cite{Liao:2017uzy,Papoulias:2017qdn,Papoulias:2018uzy,Papoulias:2019xaw,Khan:2019cvi,Dutta:2019eml},
where the $Z'$ contribution to the weak charge in \cenns
is half of that in Eq.~\eqref{Q_ll},
must be reinterpreted by
rescaling their $Z'$ coupling $g_{Z'}$ by a factor $\sqrt{2}$.

\section{Muon $g-2$}
\label{app:gm2}

Recently,
the Fermilab Muon $g-2$ experiment~\cite{Muong-2:2021ojo}
confirmed the value of the muon anomalous magnetic moment $(g-2)_\mu$
that was measured in 2006 in the Muon E821 experiment at Brookhaven National Laboratory~\cite{Muong-2:2006rrc},
leading to the combined $4.2\sigma$ deviation from the Standard Model prediction
\begin{equation}\label{eq:d_amu_exp}
\Delta a_\mu = (25.1\pm 5.9) \times 10^{-10},
\end{equation}
where $a_\mu=(g-2)_\mu/2$.
This $(g-2)_\mu$ anomaly may be due to new physics beyond the SM
(see the reviews in
Refs.~\cite{Jegerlehner:2009ry,Keshavarzi:2021eqa,Li:2021bbf}).

In theories beyond the SM,
an additional neutral boson $B$ with mass $M_B$,
which interacts with muons with coupling $g_B$,
contributes to the muon anomalous magnetic moment with~\cite{Brodsky:1967sr}
\begin{equation}\label{eq:d_amu_th}
\delta a_\mu^{B}
=
\frac{g_B^2}{8\pi^2}
\int_{0}^{1} dx
\,
\frac{ Q(x) }{ x^2 + \left( 1 - x \right) M_B^2 / m_\mu^2 }
\end{equation}
where $Q(x)$ depends on the
scalar or vector nature of the neutral boson $B$:
\begin{equation}
Q(x)
=
\left\{
\begin{array}{rl} \displaystyle
x^2 \left( 2 - x \right) & \quad \text{(scalar)}
,
\\ \displaystyle
2 x^2 \left( 1 - x \right) & \quad \text{(vector)}
.
\end{array}
\right.
\label{Q}
\end{equation}


\input{main.bbl}

\end{document}

%% file: main.bbl
%